\documentclass[
  aps,
  prb,
  reprint,
  notitlepage,
  floatfix,
  nofootinbib,
  longbibliography
]{revtex4-2}

\usepackage{graphicx}
\usepackage{booktabs}
\usepackage{amsmath,amssymb}
\usepackage{fontspec}
\usepackage{unicode-math}
\usepackage{microtype}
\usepackage{xcolor}
\usepackage{physics}
\usepackage{tikz}
\usetikzlibrary{positioning,arrows.meta,decorations.pathmorphing,
                decorations.markings,calc,shapes.geometric,backgrounds,fit,shadows}
\usepackage{hyperref}

\graphicspath{{figures/}}
\newcommand{\Displacement}{0.375\,nm}
\newcommand{\PermanentDipoleRatio}{0.72}
\newcommand{\TransitionDipole}{25\,D}
\newcommand{\TransitionDipoleLifetime}{653\,ps}
\newcommand{\ReferenceRefractiveIndex}{3.5}
\newcommand{\AcousticBandMaximum}{27.9\,meV}
\newcommand{\PublishedDephasing}{94\,meV}
\newcommand{\DephasingSweepFactor}{100}
\newcommand{\IdlerBranchingPercent}{25.8\%}
\newcommand{\DriveRabiRatio}{0.1}
\newcommand{\PlasmonBasisSize}{6}
\newcommand{\PlasmonLinewidth}{75\,meV}
\newcommand{\EmitterRadiativeLinewidth}{1.01\times10^{-3}\,meV}
\newcommand{\AntennaFraction}{0.5}
\newcommand{\IdlerWavelength}{1475\,nm}
\newcommand{\SignalWavelength}{1152\,nm}
\newcommand{\FluorescenceWavelength}{647\,nm}
\newcommand{\SymmetricSignalWavelength}{1160\,nm}
\newcommand{\SymmetricFluorescenceWavelength}{637\,nm}
\newcommand{\SymmetricLPEnergyRatio}{0.5488130144}
\newcommand{\ReferenceLPEnergyRatio}{0.552681725}
\newcommand{\ReferenceUPEnergyRatio}{0.9842343846}
\newcommand{\FluorescenceElasticFraction}{99\%}
\newcommand{\SecularLinewidthRatio}{2.8}
\newcommand{\CutoffRateChange}{0.07\%}
\newcommand{\SharedModeBound}{1.5\%}
\newcommand{\DiscardedStatePopulation}{1.9\times10^{-5}}

\newcommand{\IdlerRatePerSecond}{3.1\times10^{10}\,s^{-1}}
\newcommand{\FluorescenceRatePerSecond}{2.9\times10^{10}\,s^{-1}}
\newcommand{\FluorescenceCurrentChange}{76.5\%}
\newcommand{\RelativeCoupling}{0.275}
\newcommand{\ModeEnergy}{1.4\,eV}
\newcommand{\ExcitonEnergy}{1.62\,eV}
\newcommand{\UltrastrongThreshold}{0.1}
\newcommand{\UltrastrongPerturbativeLimit}{0.3}
\newcommand{\AttainableCoupling}{0.05}
\newcommand{\AttainableCouplingBranching}{0.47\%}
\newcommand{\ImpliedModeVolume}{1.6\,nm^{3}}
\newcommand{\PublishedModeVolume}{73\,nm^{3}}
\newcommand{\PublishedModeVolumeCoupling}{0.041}
\newcommand{\DotBaseDiameter}{40\,nm}
\newcommand{\DotHeight}{8\,nm}
\newcommand{\DotCoupling}{0.0035}
\newcommand{\ReferenceOverDotCoupling}{79}
\newcommand{\DotGapSplitting}{200\pm45\,meV}
\newcommand{\SplittingOverReported}{4.2}

\newcommand{\OffResonantFilteredPercent}{0.002\%}

\newcommand{\OpticalPairYieldPercent}{12.9\%}
\newcommand{\OpticalSignalEscapePercent}{50\%}
\newcommand{\ReferenceDriveEnergy}{1.65\,meV}
\newcommand{\ReferenceIdlerDecayEnergy}{17.6\,meV}
\newcommand{\ReferenceFluorescenceDecayEnergy}{16.5\,meV}
\newcommand{\ReferenceSignalDecayEnergy}{67.2\,meV}
\newcommand{\SlowSignalDecayEnergy}{0.672\,meV}
\newcommand{\IntermittentSignalDecayEnergy}{0.00672\,meV}
\newcommand{\IntermittentCascadeFano}{0.879}
\newcommand{\IntermittentFluorescenceFano}{1.69}
\newcommand{\IntermittentLPOccupancy}{85.9\%}
\newcommand{\DetectorResponseFwhm}{20\,ps}
\newcommand{\CoincidenceWindowHalfWidth}{20\,ps}
\newcommand{\CoincidenceToAccidentalRatio}{1.39}
\newcommand{\ConditionalSignalExcess}{0.487}
\newcommand{\OrderingLinearDelay}{5\,fs}
\newcommand{\LPLifetime}{9.79\,fs}
\newcommand{\BranchMapEscapeEnergy}{34.1\,meV}
\newcommand{\BranchMapReferenceMinimum}{0.69}
\newcommand{\BranchMapOptimalDriveEnergy}{24.1\,meV}

\definecolor{linknavy}{HTML}{1A4D8F}
\definecolor{unobservedgray}{HTML}{8A8A8A}
\definecolor{idlerC}{HTML}{7FB6DD}
\definecolor{signalC}{HTML}{93C9A0}
\definecolor{fluorC}{HTML}{F4B266}
\colorlet{idlerT}{idlerC!72!black}
\colorlet{signalT}{signalC!70!black}
\colorlet{fluorT}{fluorC!78!black}
\tikzset{
  every picture/.style={line cap=round, line join=round},
  lsh/.style={preaction={draw, -, line cap=round, line join=round, line width=#1,
              black, opacity=0.16, transform canvas={xshift=1.3pt, yshift=-1.3pt}}},
  lsh/.default=3pt,
  lvl/.style={line width=2pt, lsh=2.4pt}
}
\hypersetup{
  colorlinks=true,
  citecolor=linknavy,
  linkcolor=linknavy,
  urlcolor=linknavy,
  breaklinks=true,
  pdftitle={A semiconductor photon-pair source based on a polariton cascade},
  pdfauthor={Karol Kawa}
}

\newcommand{\Diss}[1]{\mathcal{D}[#1]}
\newcommand{\gtwo}{g^{(2)}}
\newcommand{\w}{\omega}

\begin{document}

\title{A semiconductor photon-pair source based on a polariton cascade}

\author{Karol Kawa}
\email[Contact author: ]{kawa@fzu.cz}
\affiliation{FZU --- Institute of Physics of the Czech Academy of Sciences,
Na Slovance 1999/2, 182 00 Prague 8, Czech Republic}
\date{October 20, 2026}

\begin{abstract}
We present a theoretical study of a semiconductor photon-pair source based on a radiative cascade enabled by a permanent exciton dipole.
The source consists of a GaAs quantum dot placed between a metal nanoparticle and a mirror.
The dot exciton and the localized electromagnetic mode mix to form upper and lower polaritons.
Without a permanent dipole, a symmetry separating states with even and odd excitation numbers forbids photon emission between the polaritons.
Separation of the mean electron and hole positions gives the exciton a permanent dipole and breaks this symmetry.
The upper polariton can then emit an idler photon as it decays to the lower polariton.
The lower polariton can emit a signal photon as the system returns to its ground state.
Using an effective model, we calculate the probability that a prepared upper polariton emits both photons through the metal antenna.
We also analyze fluctuations in the emitted photon counts and correlations between the two emission channels.
The calculation establishes an operating principle for assumed values of the coupling strength, emitter properties and optical loss rates.
Whether the device can be built and its photon correlations measured remains open.
\end{abstract}

\maketitle

\section{Introduction}\label{sec:introduction}

The ability to generate photon pairs is important for quantum communication and fundamental tests of quantum mechanics.
Entangled photon pairs enable tests of Bell inequalities~\cite{Freedman1972} and entanglement-based quantum-key distribution~\cite{Ekert1991,Xu2020,Yin2020}.
Photon-pair sources also allow heralded single-photon preparation, where detection of one photon indicates the presence of its partner~\cite{Zhao2020}.
These uses require properties beyond pair emission alone, such as entanglement or suppression of unwanted additional photons.
These requirements motivate the development of compact sources and the study of pair-emission mechanisms in semiconductor structures.

Two established routes to photon-pair production are spontaneous parametric down-conversion (SPDC) and radiative cascades.
In SPDC, a nonlinear optical interaction converts a pump photon into two lower-energy photons~\cite{Burnham1970}.
Such sources use nonlinear crystals~\cite{Kwiat1995} or thin-film lithium niobate waveguides~\cite{Zhao2020}.
In a radiative cascade, an excited system emits two photons in successive transitions through a real intermediate state.
Examples include atomic cascades in calcium~\cite{Freedman1972} and the biexciton-exciton cascade in semiconductor quantum dots, demonstrated with entangled pairs~\cite{Benson2000,Akopian2006}.
In the quantum-dot cascade, a biexciton containing two electron--hole pairs decays to an exciton containing one pair, which then decays to the empty dot.
This sequence supports semiconductor sources of entangled photon pairs~\cite{Liu2019,Wang2019}.
Cavity-enhanced pair emission has also been demonstrated in a regime where the biexciton-exciton cascade coexists with cavity-stimulated two-photon emission~\cite{Wu2026}.

Plasmonic cavities provide another setting for photon-pair generation by confining an electromagnetic mode near a metal structure.
One quantum of this mode is a plasmon.
Coupling an exciton to an electromagnetic mode mixes matter and field excitations into polariton states~\cite{Hopfield1958}.
Piryatinski and Sukharev~\cite{ChiSquared2023} studied photon-pair generation in an incoherently pumped ensemble of emitters coupled to a plasmonic mode with a second-order nonlinear response.
Their interaction converts one emitter excitation into two plasmons and mixes single-exciton states with two-plasmon states.
This mechanism produces degenerate parametric down-conversion, with both photons emitted in the same frequency band.

A different route uses a cascade through upper and lower polariton states.
Pompe \textit{et al.}~\cite{Pompe2023} proposed such a source using a quantum dot coupled to the localized plasmon mode between a metal nanoparticle and a mirror.
In the proposed sequence, excitation of the upper polariton is followed by a transition to the lower polariton and a return to the ground state.
They assigned a central role to exciton dephasing, the loss of phase coherence of the emitter.
For this sequence to produce a photon pair, both decay steps must emit a photon.
This raises the question of which coupling makes the upper-to-lower polariton transition radiative.

A permanent emitter dipole provides such a coupling.
Scala \textit{et al.}~\cite{Scala2021} showed that the interaction of permanent atomic dipoles with the cavity field permits photon emission between dressed states of the same excitation manifold.
Without the permanent-dipole term, the quantum Rabi Hamiltonian conserves excitation-number parity, which separates states with even and odd total numbers of emitter and mode excitations.
The upper and lower polaritons considered here have the same parity, while the electric-dipole emission operators change parity.
Photon emission between these states is therefore forbidden.
The permanent-dipole term breaks parity and permits the transition.
Separation of the mean electron and hole positions can give a quantum-dot exciton the required permanent dipole~\cite{Fry2000}.

\begin{figure*}[!t]
\centering
\includegraphics{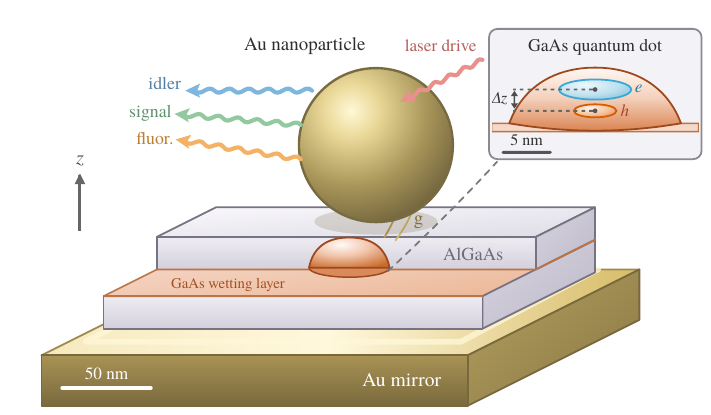}
\caption{Hypothetical device geometry.
A GaAs quantum dot lies in AlGaAs between a gold nanoparticle and a gold mirror, above a GaAs wetting layer.
The solid AlGaAs block represents a bulk matrix.
The gap axis $z$ marks the growth direction and mode polarization, and $g$ labels the exciton-mode coupling.
The red arrow marks the laser drive, and the blue, green and orange arrows mark idler, signal and fluorescence emission.
The inset shows the electron and hole densities separated by $\Delta z$.
The scale bars apply to the main drawing and inset, while the carrier ordering, mirror thickness and separation are schematic.}\label{fig:system}
\end{figure*}

Here we study the radiative polariton cascade enabled by a permanent exciton dipole in the nanoparticle-mirror source setting proposed by Pompe \textit{et al.}.
Figure~\ref{fig:system} shows the proposed device.
We select a light-hole exciton transition with an electric dipole along the growth axis $z$~\cite{Huo2014}.
We assume a separation $\Delta z$ between the mean electron and hole positions along $z$, producing a permanent exciton dipole along the same axis.
A laser excites the upper polariton, which can emit a photon as it decays to the lower polariton.
The lower polariton can then emit another photon as the system returns to its ground state.
Following the channel labels used by Pompe \textit{et al.}, we call the upper-to-lower polariton photon the idler and the lower-polariton-to-ground photon the signal.
We refer to direct upper-polariton emission to the ground state as fluorescence.

We calculate the probability that a prepared upper polariton emits both cascade photons through the metal antenna.
We include the competing fluorescence transition and the possibility that either cascade step does not produce an antenna photon.
For continuous laser excitation, we retain the two polaritons and the ground state in an effective model and apply photon-counting theory~\cite{Emary2007,Flindt2008,Marcos2010,Landi2024}.
We calculate emitted-light spectra and correlations between photon counts in the two channels.
The time-ordered correlation identifies the sequence of idler and signal emission within the model.
We also examine how detector timing affects the predicted coincidence signal and how much off-resonant drive background passes a signal filter.
The calculation establishes an operating principle for assumed values of the coupling strength, emitter properties and optical loss rates.
Whether the device can be built and its photon correlations measured remains open.

The organization of the paper is as follows.
Section~\ref{sec:system} introduces the system, the parity selection rule, and the permanent-dipole mechanism.
Section~\ref{sec:methods} defines the dynamics and photon observables.
Section~\ref{sec:results} presents the calculated emission and counting statistics.
Sections~\ref{sec:discussion} and~\ref{sec:conclusions} discuss the limitations and state the conclusions.

\section{Model system}\label{sec:system}

The total Hamiltonian of the optically driven quantum dot, the plasmon mode and their reservoirs is
\begin{align}
  \hat{H}(t) = \hat{H}_{\mathrm{S}} + \hat{H}_{\mathrm{B}} + \hat{H}_{\mathrm{I}} + \hat{H}_{\mathrm{laser}}(t),
  \label{eq:total-hamiltonian}
\end{align}
where $\hat{H}_{\mathrm{S}}$ describes the quantum dot, the localized plasmon mode and the interaction between them.
The reservoirs of radiation and absorption are described by $\hat{H}_{\mathrm{B}}$, and their coupling to the quantum dot and plasmon mode by $\hat{H}_{\mathrm{I}}$.
The term $\hat{H}_{\mathrm{laser}}(t)$ couples the plasmon mode to the classical laser field.

\subsection{Quantum dot and plasmon mode}

\subsubsection{Device and Hamiltonian}

The proposed device contains a GaAs quantum dot serving as a two-level quantum emitter, embedded in an AlGaAs matrix between a gold nanoparticle and a gold mirror (Fig.~\ref{fig:system}).
The gap axis $z$ coincides with the growth direction, along which the electric field of the retained localized surface-plasmon mode is polarized~\cite{Chikkaraddy2016}.
The emitter ground state $\ket{g}$ is the empty dot, while its excited state $\ket{e}$ is a bright light-hole exciton.
Of its fine-structure transitions, we retain the one whose dipole lies along the growth axis $z$, which couples to this mode.

A bright transition with this polarization can be obtained from a predominantly light-hole exciton under tensile strain~\cite{Huo2014}, whereas bright transitions in the pure heavy-hole limit are polarized in the growth plane.
The light-hole exciton also has an in-plane bright doublet and a dark state, which lie below the selected growth-axis transition in the cited experiment.
We retain only the selected light-hole transition and neglect biexciton excitation.
The other fine-structure states, higher orbital excitations and relaxation into those states are also omitted.
This two-level approximation is a model assumption whose accuracy depends on the couplings and relaxation rates of the omitted states~\cite{Reiter2019}.

The interaction between the emitter and plasmon mode is described by an effective quantum Rabi Hamiltonian with a permanent-dipole term in the dipole-gauge convention~\cite{DiStefano2019,Savasta2021},
\begin{align}
\begin{aligned}
    \hat{H}_{\mathrm{S}} = {} & \hbar\w_m\hat{b}^{\dagger}\hat{b}
    +\hbar\w_e\hat{\sigma}_{+}\hat{\sigma}_{-} \\
    &+\hbar\left(\hat{b}+\hat{b}^{\dagger}\right)
     \left[g\left(\hat{\sigma}_{+}+\hat{\sigma}_{-}\right)
     +\frac{d_p}{2}\hat{\sigma}_{z}\right],
\end{aligned}
    \label{eq:system-hamiltonian}
\end{align}
where $\hbar\w_e$ is the selected exciton energy relative to the empty dot and $\hbar\w_m$ is the energy of one plasmon.
The operators $\hat{\sigma}_{+}=\dyad*{e}{g}$ and $\hat{\sigma}_{-}=\dyad*{g}{e}$ create and destroy the exciton, respectively, so $\hat{\sigma}_{+} \hat{\sigma}_{-}$ gives its population and $\hat{\sigma}_{z}=\dyad*{e}{e}-\dyad*{g}{g}$ distinguishes the two emitter states.
The operator $\hat{b}$ annihilates a plasmon in the shifted mode coordinate defined below, and $\hat{b}+\hat{b}^{\dagger}$ is the dimensionless mode quadrature to which the emitter dipole couples.
The coupling $g$ describes the interaction of the transition dipole moment with the mode field, while $d_p$ describes the coupling of the permanent-dipole difference to that field.
Both couplings are in frequency units.

We use the dimensionless ratio $\eta=2g/(\w_m+\w_e)$ to compare the coupling strength with the mean bare frequency.
This is the frequency scale used to distinguish ultrastrong coupling, where counter-rotating terms become important~\cite{FornDiaz2019,FriskKockum2019}.
The coupling comparison in Sec.~\ref{sec:currents} uses $\eta$ as its control parameter, while Eq.~\eqref{eq:system-hamiltonian} keeps the counter-rotating terms throughout.
The symmetry argument below does not require ultrastrong coupling.

To relate $g$ and $d_p$ to the emitter, we write its electric-dipole operator along $z$ as $\hat{\mu}_{\parallel}$.
Its diagonal matrix elements $\mu_{gg}=\mel*{g}{\hat{\mu}_{\parallel}}{g}$ and $\mu_{ee}=\mel*{e}{\hat{\mu}_{\parallel}}{e}$ are the permanent dipole moments of the two states, while $\mu_{eg}=|\mel*{e}{\hat{\mu}_{\parallel}}{g}|$ is the transition-dipole magnitude.
We choose the state phases to make the transition matrix element positive and orient $z$ so that the permanent-dipole difference $\Delta\mu=\mu_{ee}-\mu_{gg}$ is nonnegative.
For the two-state dipole operator $\hat{\mu}_{\parallel}^{(2)}$, subtracting the common diagonal contribution gives
\begin{align}
\begin{aligned}
    \hat{\mu}_{\parallel}^{\mathrm{c}}
    &=\hat{\mu}_{\parallel}^{(2)}
      -\frac{\mu_{ee}+\mu_{gg}}{2}\hat{\mathbb{1}}\\
    &=\mu_{eg}\left(\hat{\sigma}_{+}+\hat{\sigma}_{-}\right)+\frac{\Delta\mu}{2}\hat{\sigma}_{z}.
\end{aligned}
    \label{eq:centered-dipole}
\end{align}
This centered dipole contains both the transition moment $\mu_{eg}$ and the permanent-dipole difference $\Delta\mu$.

The two dipole contributions couple to the same local electric field along $z$.
If its vacuum amplitude is $\mathcal{F}_{\mathrm{vac}}$, the couplings satisfy $\hbar g=\mathcal{F}_{\mathrm{vac}}\mu_{eg}$ and $\hbar d_p=\mathcal{F}_{\mathrm{vac}}\Delta\mu$~\cite{Scala2021}.
The common field amplitude therefore cancels from their ratio,
\begin{subequations}
\begin{align}
    \frac{d_p}{g} & = \frac{\Delta\mu}{\mu_{eg}},\\
    \Delta\mu & = e \Delta z,
\end{align}\label{eq:dipole-geometry}
\end{subequations}
where $e>0$ is the elementary charge and $\Delta z\geq0$, the electron--hole displacement, is the separation between the mean electron and hole positions along $z$.
The added exciton dipole points from the mean electron position to the mean hole position.
The ratio $d_p/g$ in Eq.~\eqref{eq:dipole-geometry}, which we call the permanent-dipole asymmetry, is defined for $g>0$ and $\mu_{eg}>0$.

The centered form is convenient because the average dipole moment can be absorbed into a shift of the oscillator coordinate.
If $\hat{c}$ annihilates a plasmon before this shift, the operator in Eq.~\eqref{eq:system-hamiltonian} is $\hat{b}=\hat{c}+\mathcal{F}_{\mathrm{vac}}(\mu_{ee}+\mu_{gg})/(2\hbar\w_m)$.
This change of coordinates leaves the spectrum of the emitter-mode Hamiltonian unchanged and is made before the plasmon number states are truncated.
The dipole-gauge Hamiltonian includes a term quadratic in the dipole, called the dipole self-energy.
In the centered form, the remaining self-energy is the constant $\hbar(g^2+d_p^2/4)/\w_m$ within the two-level model.
It shifts every level equally and can therefore be omitted without changing transition energies, giving Eq.~\eqref{eq:system-hamiltonian}.

Asymmetric confinement of the electron and hole can produce a nonzero $\Delta z$ through the dot shape, composition or strain (Fig.~\ref{fig:system}, inset).
Stark spectroscopy provides evidence for permanent exciton dipoles: in a fit of the transition energy versus an applied static electric field, the linear part is attributed to the permanent dipole and the quadratic part to polarizability~\cite{Fry2000,Huang2021}.
For weakly confined GaAs dots, the inferred dipole depends on the fitted field interval and need not give the zero-field electron--hole separation~\cite{Huang2021}.
These measurements motivate a nonzero permanent dipole in the model, but do not determine its value for the selected light-hole emitter.
An applied field would also change the exciton energy and transition dipole, so it is not a way to tune $d_p$ alone in a device.

\begin{figure*}[!t]
\centering
\begin{tikzpicture}[
  font=\normalsize,
  >={Latex[round,length=6pt,width=5pt]},
  radiative/.style={line width=1.8pt,-{Latex[round,length=6pt,width=5pt]}},
  blocked/.style={draw=unobservedgray,line width=1.25pt},
  blockedmark/.style={draw=red!85!black,line width=1.35pt},
]
\path (0,0) rectangle (13.45,6.0);
\pgfmathsetmacro{\yGS}{0.72}
\pgfmathsetmacro{\levelSpan}{4.10}

\begin{scope}
  \pgfmathsetmacro{\yLP}{\yGS+\levelSpan*\SymmetricLPEnergyRatio}
  \pgfmathsetmacro{\yUP}{\yGS+\levelSpan}
  \node[font=\bfseries,anchor=west] at (0.05,5.65) {(a)};
  \node[font=\bfseries] at (3.35,5.65) {$d_p=0$: parity conserved};
  \draw[lvl] (2.40,\yGS) -- (4.80,\yGS);
  \node[anchor=east] at (2.30,\yGS) {$\ket{\mathrm{GS}}$};
  \draw[lvl] (1.20,\yLP) -- (3.60,\yLP);
  \node[anchor=east] at (1.10,\yLP) {$\ket{\mathrm{LP}}$};
  \draw[lvl] (2.40,\yUP) -- (4.80,\yUP);
  \node[anchor=east] at (2.30,\yUP) {$\ket{\mathrm{UP}}$};
  \node[anchor=south,font=\small] at (3.60,{\yGS+0.10}) {$\Pi=+1$};
  \node[anchor=south,font=\small] at (2.40,{\yLP+0.10}) {$\Pi=-1$};
  \node[anchor=south,font=\small] at (3.60,{\yUP+0.10}) {$\Pi=-1$};

  \draw[blocked] (2.50,{\yUP-0.12}) -- (1.32,{\yLP+0.12});
  \draw[blockedmark]
    (1.82,{(\yUP+\yLP)/2-0.09}) -- (2.00,{(\yUP+\yLP)/2+0.09})
    (1.82,{(\yUP+\yLP)/2+0.09}) -- (2.00,{(\yUP+\yLP)/2-0.09});
  \draw[radiative,signalT] (1.32,{\yLP-0.12}) -- (2.50,{\yGS+0.12});
  \node[anchor=west,align=left] at (2.16,{(\yLP+\yGS)/2})
    {signal\\ $\mathrm{\SymmetricSignalWavelength}$};
  \draw[radiative,fluorT] (4.55,{\yUP-0.12}) -- (4.55,{\yGS+0.12});
  \node[anchor=west,align=left] at (4.74,{(\yUP+\yGS)/2})
    {fluorescence\\ $\mathrm{\SymmetricFluorescenceWavelength}$};
\end{scope}

\begin{scope}[shift={(6.8,0)}]
  \pgfmathsetmacro{\yLP}{\yGS+\levelSpan*\ReferenceLPEnergyRatio}
  \pgfmathsetmacro{\yUP}{\yGS+\levelSpan*\ReferenceUPEnergyRatio}
  \node[font=\bfseries,anchor=west] at (0.05,5.65) {(b)};
  \node[font=\bfseries] at (3.35,5.65) {$d_p/g=\PermanentDipoleRatio$: parity broken};
  \draw[lvl] (2.40,\yGS) -- (4.80,\yGS);
  \node[anchor=east] at (2.30,\yGS) {$\ket{\mathrm{GS}}$};
  \draw[lvl] (1.20,\yLP) -- (3.60,\yLP);
  \node[anchor=east] at (1.10,\yLP) {$\ket{\mathrm{LP}}$};
  \draw[lvl] (2.40,\yUP) -- (4.80,\yUP);
  \node[anchor=east] at (2.30,\yUP) {$\ket{\mathrm{UP}}$};

  \draw[radiative,idlerT] (2.50,{\yUP-0.12}) -- (1.32,{\yLP+0.12});
  \node[anchor=west,align=left] at (2.16,{(\yUP+\yLP)/2})
    {idler\\ $\mathrm{\IdlerWavelength}$};
  \draw[radiative,signalT] (1.32,{\yLP-0.12}) -- (2.50,{\yGS+0.12});
  \node[anchor=west,align=left] at (2.16,{(\yLP+\yGS)/2})
    {signal\\ $\mathrm{\SignalWavelength}$};
  \draw[radiative,fluorT] (4.55,{\yUP-0.12}) -- (4.55,{\yGS+0.12});
  \node[anchor=west,align=left] at (4.74,{(\yUP+\yGS)/2})
    {fluorescence\\ $\mathrm{\FluorescenceWavelength}$};
\end{scope}
\end{tikzpicture}
\caption{Radiative transitions between the three retained eigenstates of Eq.~\eqref{eq:system-hamiltonian}.
The labels give representative vacuum wavelengths for the chosen model parameters $\hbar\w_m=\mathrm{\ModeEnergy}$, $\hbar\w_e=\mathrm{\ExcitonEnergy}$ and $\eta=2g/(\w_m+\w_e)=\RelativeCoupling$.
(a) At $d_p=0$, the crossed grey line marks the forbidden idler transition, and the labels give the state parities.
(b) At $d_p/g=\PermanentDipoleRatio$, the blue arrow marks the allowed idler transition.
Green and orange arrows mark signal and fluorescence emission in both panels.
The levels are drawn relative to GS, using the same energy scale in both panels.
}\label{fig:jumps}
\end{figure*}

\subsubsection{Symmetry and parity breaking}\label{sec:selection-rule}\label{sec:permanent-dipole}

At $d_p=0$, Eq.~\eqref{eq:system-hamiltonian} reduces to the quantum Rabi model~\cite{Rabi1936,Braak2011,Xie2017}.
The rotating terms $\hat{b}\hat{\sigma}_{+}$ and $\hat{b}^{\dagger}\hat{\sigma}_{-}$ exchange one excitation between the emitter and mode, while the counter-rotating terms $\hat{b}^{\dagger}\hat{\sigma}_{+}$ and $\hat{b}\hat{\sigma}_{-}$ create or remove an exciton and a plasmon together.
These terms change the total excitation number by an even integer and conserve the parity operator
\begin{align}
    \hat{\Pi} = \exp\left[i\pi\left(\hat{b}^{\dagger}\hat{b}+\hat{\sigma}_{+}\hat{\sigma}_{-}\right)\right].
\end{align}
For the product states $\ket*{n,g}=\ket*{n}\otimes\ket*{g}$ and $\ket*{n,e}=\ket*{n}\otimes\ket*{e}$, where $n$ counts plasmons, the total excitation numbers are $n$ and $n+1$, respectively, giving
\begin{align}
\begin{aligned}
    \hat{\Pi} \ket*{n,g} & = (-1)^{n}   \ket*{n,g},\\
    \hat{\Pi} \ket*{n,e} & = (-1)^{n+1} \ket*{n,e}.
\end{aligned}
    \label{eq:product-state-parity}
\end{align}
The even sector contains $\ket*{0,g},\ket*{1,e},\ket*{2,g},\ldots$, while the odd sector contains $\ket*{0,e},\ket*{1,g},\ket*{2,e},\ldots$.
For the parameters considered here, the ground state $\ket{\mathrm{GS}}$ is the lowest even state, and the lower and upper polaritons, $\ket{\mathrm{LP}}$ and $\ket{\mathrm{UP}}$, are the two lowest odd states.
The polaritons are mainly mixtures of an exciton with no plasmon and one plasmon with the emitter in its ground state, while counter-rotating coupling also admixes an exciton-plasmon pair into GS.
Their parities are
\begin{align}
\begin{aligned}
    \hat{\Pi}\ket{\mathrm{GS}} & =  \ket{\mathrm{GS}},\\
    \hat{\Pi}\ket{\mathrm{LP}} & = -\ket{\mathrm{LP}},\\
    \hat{\Pi}\ket{\mathrm{UP}} & = -\ket{\mathrm{UP}}.
\end{aligned}
    \label{eq:dressed-state-parity}
\end{align}

The emission operators $\hat{b}+\hat{b}^{\dagger}$ and $\hat{\sigma}_{+}+\hat{\sigma}_{-}$ each change the total excitation number by one and therefore connect states of opposite parity.
For either operator, denoted by $\hat{X}$, this means $\hat{\Pi}\hat{X}\hat{\Pi}^{\dagger}=-\hat{X}$.
Since LP and UP have the same parity,
\begin{align}
 \mel{\mathrm{LP}}{\hat{X}}{\mathrm{UP}}
 =-\mel{\mathrm{LP}}{\hat{X}}{\mathrm{UP}}=0,
\end{align}
which forbids the radiative idler transition in Fig.~\ref{fig:jumps}(a).
Coupling to a photon reservoir cannot change this result, because the emission amplitude contains this zero system matrix element.

The permanent-dipole term changes the plasmon number by one without changing the emitter population and is parity odd,
\begin{align}
    \hat{\Pi}\left(\hat{b} + \hat{b}^{\dagger}\right) \hat{\sigma}_{z} \hat{\Pi}^{\dagger}
    = -\left(\hat{b} + \hat{b}^{\dagger}\right) \hat{\sigma}_{z}.
\end{align}
A nonzero $d_p$ therefore mixes the even and odd sectors and permits the radiative idler transition~\cite{Braak2011,Scala2021}, as shown in Fig.~\ref{fig:jumps}(b).
For nondegenerate eigenstates of the parity-conserving Hamiltonian at $d_p=0$, first-order perturbation theory gives a leading idler amplitude proportional to $d_p$.
Because the rate contains the squared amplitude, it is quadratic in $\Delta z$ near zero separation when this linear contribution is nonzero and the reservoir spectral weight is smooth and nonzero at the limiting idler energy.
This statement holds at fixed $g$, $\mu_{eg}$, bare frequencies and reservoir spectra.
The finite-displacement calculations in Sec.~\ref{sec:currents} use diagonalized states and recompute every transition rate at fixed reservoir weights.

The exciton population operator $\hat{\sigma}_{+}\hat{\sigma}_{-}$ is parity even and can transfer population between LP and UP through a noise reservoir.
Such population transfer is not an idler photon and does not open the forbidden electromagnetic transition at $d_p=0$.
The population-noise coupling and the full emitter radiation operator are distinguished in Sec.~\ref{sec:reservoirs}.

The finite-displacement reference states are the three distinct eigenstates that maximize their total squared overlap with GS, LP and UP at $d_p=0$.
For the asymmetry and coupling scans, we follow the eigenstate branches through this reference point by maximizing the total squared overlap between distinct states at adjacent scan points.
The asymmetry scan starts at zero permanent dipole with the reference coupling, while the coupling scan keeps the reference displacement fixed.
At stronger coupling, the continued UP need not be the eigenstate with the largest overlap with UP at $d_p=0$.
The statement that the idler vanishes at $d_p=0$ along the coupling scan instead uses parity labels at each coupling, with LP and UP the two lowest odd states.
The open-system calculation retains these three dressed states, which is a separate approximation from restricting the emitter to two states or truncating the plasmon number states.
Section~\ref{sec:generator} describes the truncation and omitted-drive checks.

\subsection{Reservoirs}\label{sec:reservoirs}

Energy leaves the coupled emitter and plasmon mode through antenna radiation, absorption in the metal and direct emitter radiation.
We represent these processes by external reservoirs and assume weak system-reservoir coupling and short reservoir memory, as required for the Born--Markov treatment in Sec.~\ref{sec:rates}.
A localized lossy mode coupled to external continua can be obtained from electromagnetic quantization~\cite{Franke2019}, but identifying that mode and separating it from the remaining response of this nanogap is an assumption of the present effective model~\cite{Gustin2023,GustinErratum2025}.

The gold nanoparticle and mirror form the antenna, whose radiation reservoir is labeled $\mathrm{R}$, while $\mathrm{A}$ labels absorption in the metal.
The harmonic modes used for absorption represent the metal's dissipative response~\cite{Dung1998,Franke2019}.
Direct emitter radiation into the remaining electromagnetic background is labeled $\mathrm{E}$.
For each reservoir $\alpha\in\{\mathrm{R},\mathrm{A},\mathrm{E}\}$, the bosonic operator $\hat{a}_{\alpha\ell}$ annihilates an excitation of mode $\ell$ with frequency $\w_{\alpha\ell}$, giving the free reservoir Hamiltonian
\begin{align}
  \hat{H}_{\mathrm{B}} = \sum_{\alpha,\ell}
  \hbar\w_{\alpha\ell} \hat{a}_{\alpha\ell}^{\dagger}\hat{a}_{\alpha\ell}.
\end{align}
These operators obey $[\hat{a}_{\alpha\ell},\hat{a}_{\beta n}^{\dagger}]=\delta_{\alpha\beta}\delta_{\ell n}$ and commute with the emitter and localized-mode operators.

Both antenna radiation and metal absorption couple through the mode quadrature $\hat{b}+\hat{b}^{\dagger}$.
Direct emitter radiation couples through the full centered dipole, normalized by the transition moment,
\begin{align}
  \frac{\hat{\mu}_{\parallel}^{\mathrm{c}}}{\mu_{eg}}
  = \hat{\sigma}_{+} + \hat{\sigma}_{-} + \frac{d_p}{2g} \hat{\sigma}_{z}.
  \label{eq:emitter-radiative-operator}
\end{align}
The transition and permanent-dipole terms belong to the same radiation operator and couple to the same photon reservoir.
The interaction with all three reservoirs is therefore
\begin{align}
\begin{aligned}
  \hat{H}_{\mathrm{I}}
  ={} & \left(\hat{b} + \hat{b}^{\dagger} \right)
        \sum_{\alpha \in \{\mathrm{R},\mathrm{A}\}}
        \sum_{\ell}
        \left(
        v_{\alpha\ell} \hat{a}_{\alpha\ell} + v_{\alpha\ell}^{*} \hat{a}_{\alpha\ell}^{\dagger}
        \right) \\
      & + \left( \hat{\sigma}_{+} + \hat{\sigma}_{-}
        + \frac{d_p}{2g} \hat{\sigma}_{z} \right)\\
  &\quad\times\sum_{\ell}
    \left(
    v_{\mathrm{E}\ell} \hat{a}_{\mathrm{E}\ell}
    +v_{\mathrm{E}\ell}^{*} \hat{a}_{\mathrm{E}\ell}^{\dagger}
    \right),
\end{aligned}
\label{eq:reservoir-hamiltonian}
\end{align}
where $v_{\alpha\ell}$ are coupling amplitudes in energy units, with the complex-conjugate terms ensuring that $\hat{H}_{\mathrm{I}}$ is Hermitian.
This convention matches the energy spectral densities defined below in Eq.~\eqref{eq:spectral-density}.

For distinct dressed states, $\mel{j}{\hat{\sigma}_{z}}{k}=2\mel{j}{\hat{\sigma}_{+}\hat{\sigma}_{-}}{k}$, so the direct-emitter idler amplitude contains both the transition-dipole matrix element and $(d_p/g)\mel{\mathrm{LP}}{\hat{\sigma}_{+}\hat{\sigma}_{-}}{\mathrm{UP}}$.
The rate is proportional to the squared modulus of their sum and includes interference between these contributions.
Both the emitter radiation operator and the dressed states change with $d_p$.
At $d_p=0$, the radiative idler matrix element vanishes by parity even though the exciton population matrix element can be nonzero.

We treat the three reservoirs as independent.
For the two radiation reservoirs, this assumes that the photon modes assigned to antenna radiation and direct emitter radiation are separate.
Shared photon modes would instead produce interference between the two radiation paths.
For photon modes $\ell$ coupled to both paths with amplitudes $v_{\mathrm{R}\ell}$ and $v_{\mathrm{E}\ell}$, each transition rate in Eq.~\eqref{eq:photon-jumps} below would gain a cross term between the antenna and emitter amplitudes, proportional to the cross spectral density $J_{\mathrm{RE}}(E)=\sum_{\ell}v_{\mathrm{R}\ell}v_{\mathrm{E}\ell}^{*}\delta(E-\hbar\w_{\ell})$.
The Cauchy--Schwarz inequality gives $|J_{\mathrm{RE}}|\le\sqrt{J_{\mathrm{R}}J_{\mathrm{E}}}$, with equality only when both paths couple to the same field pattern.
Here $J_{\mathrm{R}}$ and $J_{\mathrm{E}}$ are the spectral densities of the two radiation reservoirs, defined in Eq.~\eqref{eq:spectral-density}.
With the reference weights of Sec.~\ref{sec:rates}, the cross term would change the combined antenna and direct-emitter radiative rate of each transition by at most \SharedModeBound{}, because the emitter alone radiates much more weakly than the antenna.
The direct-emitter energy linewidth $\Gamma_{\mathrm{E}}$ is estimated from the assumed dipole in a homogeneous medium and used for the residual background.
Adding a total nanogap-enhanced emitter linewidth to the explicitly retained localized mode could count the same electromagnetic response twice~\cite{Franke2019,Gustin2023,GustinErratum2025}.
A device calculation of the mode fields, residual background, radiation and absorption would be needed to test this separation and determine whether collection can isolate antenna radiation.

The reference reservoirs are stationary, with zero mean coupling fields and negligible thermal occupation at the retained optical transition energies.
Their correlation functions therefore depend only on the time difference.
Reservoir-induced energy shifts and zero-frequency electromagnetic noise are omitted.

The mode energy linewidth $\Gamma_m$ includes radiation and absorption, with fractions $r_m$ and $1-r_m$, respectively.
Only antenna radiation contributes to the counted photon observables, while metal absorption and direct emitter radiation remain unobserved loss.
Propagation, collection and detector losses act after antenna emission and are treated separately from $r_m$.
An optional nonradiative emitter reservoir couples through $\hat{\sigma}_{+}+\hat{\sigma}_{-}$ with weight $\Gamma_{\mathrm{N}}$, which is set to zero in the reference calculation.

Fluctuations of the exciton energy couple through $\hat{\sigma}_{+}\hat{\sigma}_{-}$ and cause pure dephasing of an isolated emitter.
In the coupled emitter-mode system, the same operator can also transfer population between dressed states~\cite{Beaudoin2011}.
These transfer rates depend on the noise spectrum at the transition frequencies, while pure dephasing depends on its zero-frequency component, so a single bare-emitter dephasing linewidth does not specify the required noise spectrum.
We omit this noise from the steady-state calculation.
The dephasing stress test of Sec.~\ref{sec:rates} varies the population-noise strength and counts electromagnetic idler events separately from population transfer, checking that dephasing does not open the forbidden idler transition.

We also neglect phonon effects in the calculated dynamics.
For the retained undriven transitions, energy-conserving one-acoustic-phonon relaxation is excluded because their energies exceed the measured GaAs acoustic band maximum of $\mathrm{\AcousticBandMaximum}$~\cite{Strauch1990}.
This comparison does not constrain changes to coherences or driven dynamics, finite phonon linewidths, or multiphonon relaxation, whose omission is a separate model assumption.

\subsection{Laser drive}\label{sec:drive}

The classical laser drives the localized plasmon mode through
\begin{align}
  \hat{H}_{\mathrm{laser}}(t) = \hbar\Omega_0
  \cos(\w_{\mathrm{d}}t) \left(\hat{b} + \hat{b}^{\dagger}\right),
    \label{eq:laser-drive}
\end{align}
where $\Omega_0$ is the drive amplitude in frequency units and $\w_{\mathrm{d}}$ is the laser frequency.
We choose the laser to be resonant with the ground-to-upper-polariton transition, so $\hbar\w_{\mathrm{d}}=E_{\mathrm{UP}}-E_{\mathrm{GS}}$, where $E_j$ is an undriven eigenenergy.
The mode quadrature also couples the drive to other transitions, including the off-resonant ground-to-lower-polariton transition.

The steady-state calculation in Sec.~\ref{sec:generator} retains the resonant component and makes a rotating-wave approximation for the laser drive.
Section~\ref{sec:generator} defines the resulting Rabi frequency $\Omega_{\mathrm{R}}$.
This drive approximation leaves the counter-rotating emitter-mode coupling in $\hat{H}_{\mathrm{S}}$ intact.
It requires the drive to be slow compared with the optical oscillation and the omitted drive couplings to be small compared with their detunings.
The separate treatment of decay transitions also requires sufficiently resolved frequencies.
Section~\ref{sec:generator} gives these checks at the reference parameters, but the wider drive and rate scans do not establish the conditions throughout the scanned range.

\section{Methods}\label{sec:methods}

\subsection{Dressed basis and optical matrix elements}\label{sec:dressed-elements}

We keep only GS, LP and UP in the driven dynamics.
These are eigenstates of the undriven Hamiltonian,
\begin{align}
  \hat{H}_{\mathrm{S}}\ket{j}=E_j\ket{j},
  \qquad j\in\{\mathrm{GS},\mathrm{LP},\mathrm{UP}\}.
\end{align}
The three downward transitions connect UP to LP (idler, $i$), LP to GS (signal, $s$), and UP to GS (fluorescence, $f$), as shown in Fig.~\ref{fig:jumps}.

The loss rates depend on the matrix elements of the system coupling operators between these dressed states,
\begin{align}
\begin{aligned}
 X_{jk}^{(m)}&=\mel{j}{\hat{b}+\hat{b}^{\dagger}}{k},\\
 X_{jk}^{(\mathrm{E})}&=\mel{j}{\hat{\sigma}_{+}+\hat{\sigma}_{-}
 +\frac{d_p}{2g}\hat{\sigma}_{z}}{k},\\
 X_{jk}^{(\mathrm{N})}&=\mel{j}{\hat{\sigma}_{+}+\hat{\sigma}_{-}}{k}.
\end{aligned}
\label{eq:matrix-element}
\end{align}
The label $m$ denotes total localized-mode loss, $\mathrm{E}$ denotes direct emitter radiation, and $\mathrm{N}$ denotes optional emitter nonradiative loss.
Radiation and absorption of the localized mode have the same matrix element $X_{jk}^{(m)}$.
In the rate formulas below, the superscript $\alpha\in\{m,\mathrm{E},\mathrm{N}\}$ runs over these three loss contributions, where $m$ combines the reservoirs $\mathrm{R}$ and $\mathrm{A}$ of Sec.~\ref{sec:reservoirs}.
At $d_p=0$, parity makes $X_{\mathrm{LP},\mathrm{UP}}^{(\alpha)}=0$ for every loss contribution $\alpha$.
At $d_p\ne0$ the eigenstates mix the two parity sectors, so $X_{\mathrm{LP},\mathrm{UP}}^{(\alpha)}$ becomes nonzero.
The same radiative selection rule occurs in the Jaynes--Cummings limit~\cite{JaynesCummings1963}.
To compare the reference coupling with reported mode volumes in Sec.~\ref{sec:limitations}, we estimate the coupling from the vacuum field of a mode with effective volume $V$, $\hbar g=\mu_{eg}\sqrt{\hbar\w_m/(2\epsilon_0\epsilon_r V)}$.
Here $\epsilon_0$ is the vacuum permittivity and $\epsilon_r=n_{\mathrm{ref}}^2$ is the relative permittivity around the emitter, with the reference refractive index $n_{\mathrm{ref}}$ of Sec.~\ref{sec:results}.

\subsection{Spectral densities and phenomenological rates}\label{sec:rates}

The spectral density of each reservoir in Sec.~\ref{sec:reservoirs} sets the strength of the dissipative transitions induced by that reservoir.
We assume stationary independent reservoirs and use a weak-coupling Born--Markov description, which neglects persistent system-reservoir correlations and reservoir memory~\cite{Carmichael1993}.
The spectral density collects the couplings of Eq.~\eqref{eq:reservoir-hamiltonian} at energy $E$,
\begin{align}
 J_\alpha(E)=\sum_{\ell}|v_{\alpha\ell}|^2
 \delta(E-\hbar\w_{\alpha\ell}).
 \label{eq:spectral-density}
\end{align}
The Dirac delta selects reservoir modes at energy $E$, and $J_\alpha$ has energy units.
Independent radiation and absorption give the total localized-mode spectral density $J_m(E)=J_{\mathrm{R}}(E)+J_{\mathrm{A}}(E)$.
The optional nonradiative reservoir, $\alpha=\mathrm{N}$, couples through $\hat{\sigma}_{+}+\hat{\sigma}_{-}$ with independent coupling energies $v_{\mathrm{N}\ell}$.

The jump operators follow from second-order perturbation theory in the reservoir couplings, carried out in the eigenbasis of $\hat{H}_{\mathrm{S}}$ rather than in the uncoupled emitter and mode basis.
Each system coupling is resolved into the three transitions of Fig.~\ref{fig:jumps}.
Each transition acquires a rate set by the corresponding spectral density at the transition energy and by the squared matrix element of Eq.~\eqref{eq:matrix-element}.
Terms connecting different transition frequencies are dropped, which is the secular approximation examined at the end of Sec.~\ref{sec:generator}.
This procedure gives a global Lindblad equation, whose transitions connect eigenstates of the coupled Hamiltonian~\cite{Carmichael1993,Lindblad1976,Beaudoin2011,Settineri2018,Farina2019,Trushechkin2021}.
For a downward transition $a:k\to j$, the transition energy is $E_a=E_k-E_j>0$.
Reservoir $\alpha$ contributes the population-decay rate
\begin{align}
 \gamma_a^{(\alpha)}
 =\frac{2\pi}{\hbar}J_\alpha(E_a)
 |X_{jk}^{(\alpha)}|^2.
 \label{eq:photon-jumps}
\end{align}
The total rate is $\gamma_a=\sum_\alpha\gamma_a^{(\alpha)}$, with units of inverse time.
Once the Hamiltonian and reservoir spectra are specified, these expressions determine every transition rate.
Changing the permanent dipole changes the dressed states and transition energies, so the rates must be recalculated.
The Markov approximation constrains the reservoir correlation time and permits different spectral weights at distinct transition energies~\cite{Beaudoin2011}.

The spectral densities of the proposed nanogap have not been calculated.
We specify reference reservoir weights at the retained transition energies using the energy linewidths $\Gamma_\alpha$,
\begin{align}
 J_\alpha(E_a)=\frac{\Gamma_\alpha}{2\pi}.
 \label{eq:em-spectra}
\end{align}
For these weights, the reservoir contribution to each transition rate is $\gamma_a^{(\alpha)}=\Gamma_\alpha|X_{jk}^{(\alpha)}|^2/\hbar$.
The weights are equal at the three retained energies for each reservoir.
This choice specifies no spectral shape between those energies or outside them.
The matrix elements are recomputed at every displacement, so the reference transition rates change with displacement.

To study decay-rate dependence at fixed Hamiltonian, we vary the total rate $\gamma_a$ on transition $a\in\{i,s,f\}$ while preserving the relative contributions of its reservoirs,
\begin{align}
 \gamma_a^{(\alpha)}=\gamma_a
 \frac{\Gamma_\alpha|X_{jk}^{(\alpha)}|^2}
 {\sum_{\alpha'}\Gamma_{\alpha'}|X_{jk}^{(\alpha')}|^2}.
 \label{eq:phenomenological-rates}
\end{align}
This expression applies to allowed transitions with a nonzero denominator.
It preserves the antenna-counted fraction of each transition and the interference within the full emitter dipole.
At exact parity the idler matrix elements vanish, and the idler rate remains zero.
These rate comparisons use the fixed, nonzero reference displacement.
Choosing a different total rate represents a different assumed set of spectral weights through Eq.~\eqref{eq:photon-jumps}.
The displacement scans keep the reference weights fixed and do not assign transition rates independently of the asymmetry.
A quantitative reservoir spectrum requires the electromagnetic response, field normalization and coupling operator of the device~\cite{Gustin2025Reservoir}.
Those quantities must be consistent with the effective Hamiltonian and its gauge~\cite{Gustin2023,GustinErratum2025}.

The rate comparison of Sec.~\ref{sec:statistics} also varies the drive.
These comparisons do not establish that a device modification changes one rate independently of the energies or collection fractions.
Positive rates give a Lindblad generator, but its physical application still requires weak reservoir coupling, short reservoir memory and sufficiently separated transition frequencies~\cite{Trushechkin2021}.

Thermal upward optical transitions are neglected at these optical energies.
Zero-frequency optical noise is omitted separately.
This omission does not follow from parity, because the coupling operators can have nonzero diagonal matrix elements at finite permanent dipole.

The independence of the reservoirs and the numerical values of their transition rates require electromagnetic validation of the specific device.
These assumptions define a reference model, not an upper bound on photon emission.
We test the parity rule against dephasing in a separate pulsed calculation, the dephasing stress test.
It repeats the pulsed protocol and parameters of Ref.~\cite{Pompe2023} at $d_p=0$ and sweeps the population-coupled dephasing strength from zero to \DephasingSweepFactor{} times the largest pure-dephasing strength assumed there, $\mathrm{\PublishedDephasing}$.
The same evolved populations are counted twice.
One count assigns every UP-to-LP population transfer to the idler, and the other counts only radiative idler transitions.

\subsection{Drive projection and the master equation}\label{sec:generator}

The steady-state calculation keeps only the resonant ground-to-upper-polariton component of the drive in Eq.~\eqref{eq:laser-drive}.
In a frame rotating at the laser frequency, this component contains a constant term and a term oscillating at twice that frequency.
Dropping the oscillating term gives the drive rotating-wave approximation,
\begin{align}
  \hat{H}_{\mathrm{d}} = \frac{\hbar\Omega_{\mathrm{R}}}{2}
  \left(
      \dyad{\mathrm{UP}}{\mathrm{GS}} +
      \dyad{\mathrm{GS}}{\mathrm{UP}}
  \right).
  \label{eq:projected-drive}
\end{align}
The state phases are chosen so that the retained drive matrix element is real and positive.
The projected Rabi frequency is
$\Omega_{\mathrm{R}} = \Omega_0 \left|\mel{\mathrm{UP}}{\hat{b}+\hat{b}^{\dagger}}{\mathrm{GS}}\right|$,
where the matrix element is the one defined in Eq.~\eqref{eq:matrix-element}.
The coherent laser enters the Hamiltonian and does not produce a jump operator.

The drive is specified by $\hbar\Omega_{\mathrm{R}}$.
Each transition-rate comparison holds $\Omega_{\mathrm{R}}$ and the Hamiltonian fixed.
The laboratory drive amplitude $\Omega_0$ is therefore fixed in these comparisons.

The three downward loss jumps are
\begin{align}
\begin{aligned}
    \hat{L}_i^{\mathrm{loss}}&=\sqrt{\gamma_i}\,\dyad{\mathrm{LP}}{\mathrm{UP}},\\
    \hat{L}_s^{\mathrm{loss}}&=\sqrt{\gamma_s}\,\dyad{\mathrm{GS}}{\mathrm{LP}},\\
    \hat{L}_f^{\mathrm{loss}}&=\sqrt{\gamma_f}\,\dyad{\mathrm{GS}}{\mathrm{UP}}.
\end{aligned}
\label{eq:retained-loss-jumps}
\end{align}
Each rate sums localized-mode loss, direct emitter radiation, and any additional emitter nonradiative loss on that transition.
Figures give $\hbar\Omega_{\mathrm{R}}$ and $\hbar\gamma_a$ in energy units.
These quantities are the drive and population-decay scales, not emitted photon energies or prescribed optical spectral linewidths.
Section~\ref{sec:rates} gives the individual reservoir rates.

The rotating-frame Hamiltonian describes the coherent dynamics relative to the laser phase,
\begin{align}
\begin{aligned}
 \hat{H}_{\mathrm{rot}}
 &=\sum_j E_j\dyad{j}\\
 &\quad-\hbar\w_{\mathrm{d}}\left(\dyad{\mathrm{UP}}+\frac{1}{2}\dyad{\mathrm{LP}}\right)
 +\hat{H}_{\mathrm{d}}.
\end{aligned}
\label{eq:rotating-hamiltonian}
\end{align}
The sum over $j$ contains the three states of Sec.~\ref{sec:dressed-elements}.
The coefficient $1/2$ assigned to LP fixes the rotating-frame phase convention, rather than its physical energy.

The reduced density matrix $\hat{\rho}$ evolves according to $\dot{\hat{\rho}}=\mathcal{L}\hat{\rho}$.
The Liouvillian $\mathcal{L}$ combines this coherent evolution with the three loss channels,
\begin{align}
\begin{aligned}
 \mathcal{L}\hat{\rho}
 &=-\frac{i}{\hbar}[\hat{H}_{\mathrm{rot}},\hat{\rho}]\\
 &\quad+\sum_{a\in\{i,s,f\}}\Diss{\hat{L}_a^{\mathrm{loss}}}\hat{\rho}.
\end{aligned}
\label{eq:master-equation}
\end{align}
Here $\Diss{\hat{L}}\hat{\rho}=\hat{L}\hat{\rho}\hat{L}^{\dagger}-\{\hat{L}^{\dagger}\hat{L},\hat{\rho}\}/2$ is the Lindblad dissipator associated with a loss jump $\hat{L}$.
Each loss jump acquires only a scalar phase under this rotation, which cancels from its dissipator.
Changing the LP phase convention therefore leaves populations and photon-counting observables unchanged.

The model retains the full density matrix of GS, LP and UP, including their coherences.
The loss jumps treat transitions separately, which neglects reservoir-induced interference between different transition frequencies.
This is the secular approximation.
The neglected terms oscillate at differences between transition frequencies, so the approximation requires $|E_a-E_b|$ to be large compared with the dissipative broadening~\cite{Trushechkin2021}.
For the reference parameters, the closest transitions are the idler and signal.
Their coherence halfwidths are $\hbar(\gamma_i+\gamma_f+\gamma_s)/2$ for the idler and $\hbar\gamma_s/2$ for the signal.
Each halfwidth is the decay rate of the corresponding transition coherence multiplied by $\hbar$.
The idler--signal energy separation is $\SecularLinewidthRatio$ times the sum of these halfwidths.
This gives only moderate separation and does not establish a small error from the secular approximation in the photon-pair observables.
A quantitative error estimate would require a comparison with dynamics retaining the neglected terms.
Repeating both scans of Fig.~\ref{fig:currents} with a larger plasmon basis changes their localized-mode loss rates by at most \CutoffRateChange{}.
At the reference parameters, the largest leading-order Floquet estimate~\cite{Sambe1973} of an omitted-state population is $\DiscardedStatePopulation$.
Section~\ref{sec:offresonant} tests the omitted ground-to-lower-polariton drive within the same three-state space.

\subsection{Photon currents and branching fractions}\label{sec:counted-currents}

Only antenna radiation is counted, with antenna radiative fraction $r_m$.
Direct emitter radiation and metal absorption remain unobserved.
Let $\gamma_a^{(m)}$ be the full localized-mode loss rate on the transition that channel $a$ detects.
It is the general reservoir rate of Eq.~\eqref{eq:photon-jumps}, evaluated for the localized-mode reservoir:
\begin{align}
 \hbar\gamma_a^{(m)}=2\pi J_m(E_a)|X_{jk}^{(m)}|^2.
 \label{eq:mode-rate}
\end{align}
For the reference reservoir weights, $2\pi J_m(E_a)=\Gamma_m$.
The idler mode-loss rate therefore grows from zero through its displacement-dependent matrix element.
The three antenna-output jumps are
\begin{align}
\begin{aligned}
 \hat{L}_i&=\sqrt{r_m\gamma_i^{(m)}}\,\dyad{\mathrm{LP}}{\mathrm{UP}},\\
 \hat{L}_s&=\sqrt{r_m\gamma_s^{(m)}}\,\dyad{\mathrm{GS}}{\mathrm{LP}},\\
 \hat{L}_f&=\sqrt{r_m\gamma_f^{(m)}}\,\dyad{\mathrm{GS}}{\mathrm{UP}}.
\end{aligned}
\end{align}
For channel $a\in\{i,s,f\}$, the current superoperator retains the state change associated with its detected jump~\cite{Emary2007,Flindt2008,Marcos2010,Landi2024},
\begin{align}
  \mathcal{J}_a\hat{\rho}=\hat{L}_a\hat{\rho}\hat{L}_a^{\dagger}.
  \label{eq:current-superoperator}
\end{align}
The steady-state photon current is
\begin{align}
  I_a = \operatorname{Tr}[\mathcal{J}_a \hat{\rho}_{\mathrm{ss}}].
\end{align}
Here $\hat{\rho}_{\mathrm{ss}}$ is the normalized steady state satisfying $\mathcal{L}\hat{\rho}_{\mathrm{ss}}=0$.
The current $I_a$ is the mean number of antenna-output photons per unit time in channel $a$, before external losses.
The steady-state occupation of state $j\in\{\mathrm{GS},\mathrm{LP},\mathrm{UP}\}$ is
\begin{align}
 p_j^{\mathrm{ss}}=\mel{j}{\hat{\rho}_{\mathrm{ss}}}{j}.
 \label{eq:steady-state-occupation}
\end{align}
It gives the probability of finding the system in that state under continuous drive after transient dynamics have decayed.
The counted fraction of transition-$a$ events is $\epsilon_a=r_m\gamma_a^{(m)}/\gamma_a$.
The fraction $r_m$ refers to localized-mode loss alone, whereas $\epsilon_a$ includes every retained loss path in its denominator.
Metal absorption and emitter nonradiative loss contribute no detected photons.
For any one localized-mode loss jump $\hat{L}_m$, the identity
\begin{align}
  \Diss{\hat{L}_m}
  = \Diss{\sqrt{r_m}\hat{L}_m}
  + \Diss{\sqrt{1-r_m}\hat{L}_m}
\end{align}
shows why this detection split does not change the total localized-mode dissipator.
More generally, independent jumps proportional to the same resolved transition can be replaced by one jump whose rate is their sum.
The counted jumps are specified in addition to the loss generator, because the generator alone does not identify which outputs are observed.
No extra dissipator for a counted jump is added to Eq.~\eqref{eq:master-equation}.
Each detected channel is named for its transition, and all three emit through the localized-mode quadrature $\hat{b}+\hat{b}^{\dagger}$.
The fluorescence channel is therefore antenna radiation on the driven transition, rather than direct emitter radiation through Eq.~\eqref{eq:emitter-radiative-operator}.
In the projected steady-state model, the idler and signal have zero mean jump amplitudes, $\langle\hat{L}_i\rangle=\langle\hat{L}_s\rangle=0$, so they carry no elastic component.
The counting model treats the three transitions as separately resolved optical channels.
Finite filters can mix their spectral tails and reduce the usable currents, which requires a filter-dependent detection model.
In the electric-dipole approximation, the $z$-oriented localized mode emits no radiation along $z$, so collection must use off-axis radiation.
The antenna response determines the angular distribution available for collection.
Excluding direct-emitter photons also requires negligible emitter-field overlap with the collected output mode.
Off-axis collection, polarization, and transition energy alone do not establish that exclusion, because the two modeled dipoles share the growth-axis polarization and the dressed transition energies.
For independent loss of each antenna photon, let $\zeta_a$ be the probability that an antenna photon in channel $a$ is collected, transmitted and detected.
The recorded jump map is $\mathcal{J}_a^{\mathrm{rec}}=\zeta_a\mathcal{J}_a$.
This additional loss changes the count statistics while leaving $\mathcal{L}$ unchanged~\cite{Landi2024}.
The detector timing response used below broadens event times and does not model these efficiencies.

Two kinds of branching fraction are used below: one conditional on localized-mode loss and one for photons that leave through the antenna.
The mode branching fraction $B_m$ is the probability of taking the idler transition, conditional on an upper polariton decaying through localized-mode loss,
\begin{align}
 B_m = \frac{\gamma_i^{(m)}}{\gamma_i^{(m)}+\gamma_f^{(m)}}.
\label{eq:mode-branching}
\end{align}
The antenna-output branching fractions $B_i$ and $B_f$ are the probabilities that an upper-polariton decay emits an idler or fluorescence photon through the antenna, respectively.
The remaining probability $B_u$ describes upper-polariton decay outside these antenna channels, including absorption and direct-emitter radiation,
\begin{align}
\begin{aligned}
 B_i &= \frac{r_m\gamma_i^{(m)}}{\gamma_{\mathrm{UP}}},
 \qquad B_f = \frac{r_m\gamma_f^{(m)}}{\gamma_{\mathrm{UP}}},\\
 B_u &= 1-B_i-B_f.
\end{aligned}
\label{eq:branching-fractions}
\end{align}
Here $\gamma_{\mathrm{UP}}=\gamma_i+\gamma_f$ is the total upper-polariton loss rate.
Each total rate $\gamma_a$ sums the localized-mode, direct-emitter and nonradiative contributions $\gamma_a^{(m)}$, $\gamma_a^{(\mathrm{E})}$ and $\gamma_a^{(\mathrm{N})}$ on transition $a$.
The three mutually exclusive upper-polariton outcomes satisfy $B_i+B_f+B_u=1$.
The mode branching fraction can be recovered as $B_m=B_i/(B_i+B_f)$ because the two antenna channels share the radiative fraction $r_m$.
Signal emission starts from LP and therefore does not compete in the upper-polariton branching fractions.

For an upper polariton prepared once and allowed to decay with the laser off, the pair yield $Y_{is}$ is the probability that both cascade photons leave through the antenna,
\begin{align}
 Y_{is}=B_i B_{s|\mathrm{LP}}.
\label{eq:pair-yield}
\end{align}
An antenna idler event prepares LP, so the remaining factor is the signal escape probability $B_{s|\mathrm{LP}}$, the probability that LP emits its signal photon through the antenna,
\begin{align}
 B_{s|\mathrm{LP}}=\frac{r_m\gamma_s^{(m)}}{\gamma_s}.
\label{eq:signal-escape-probability}
\end{align}
This probability equals the counted fraction $\epsilon_s$ of signal events.
The pair yield excludes the probability of preparing UP with an incident laser pulse and excludes external collection and detector losses.
For fixed total loss rates and a common antenna radiative fraction, $Y_{is}$ is proportional to $r_m^2$.

\subsection{Counted and unobserved evolution}

For dimensionless real counting variables $\chi_a$, weighting every counted event in channel $a$ by $e^{\chi_a}$ gives the tilted generator~\cite{Landi2024},
\begin{align}
 \mathcal{L}(\boldsymbol{\chi})=\mathcal{L}
 +\sum_a(e^{\chi_a}-1)\mathcal{J}_a.
 \label{eq:tilted-generator}
\end{align}
The trace of its evolution is the moment-generating function for the photon counts.
The ordinary trace-preserving generator is recovered at $\boldsymbol{\chi}=\boldsymbol{0}$.

If all three antenna channels are monitored, evolution conditioned on no antenna count is generated by
\begin{align}
 \mathcal{L}_0=\mathcal{L}-\sum_a\mathcal{J}_a.
 \label{eq:no-count-generator}
\end{align}
Only the counted jump terms are removed.
Metal absorption and unobserved emitter jumps still change the conditioned state.
The trace of $e^{\mathcal{L}_0t}\hat{\rho}(0)$ is the probability of no monitored count up to time $t$.
Dividing that state by its trace gives the normalized conditional state.
With external inefficiency, the same construction uses $\mathcal{J}_a^{\mathrm{rec}}$ and retains missed antenna events among the unobserved jumps.

\subsection{Correlations, coincidences, and noise}\label{sec:noise}

The ordered correlation for a photon in channel $a$ followed by one in channel $b$, allowing intervening detections, is
\begin{align}
  \gtwo_{a\to b}(\tau) = \frac{\operatorname{Tr}[\mathcal{J}_b e^{\mathcal{L} \tau}
  \mathcal{J}_a\hat{\rho}_{\mathrm{ss}}]}{I_a I_b}.
  \label{eq:ordered-correlation}
\end{align}
The delay $\tau\geq0$ is measured from the detection in channel $a$ to the later detection in channel $b$.
The propagator retains every intervening event, so this correlation is not a next-event waiting-time distribution.
After any idler transition, observed or not, LP can decay only through the signal transition in this model.
That transition includes absorption and unobserved radiation, so observing an idler does not guarantee an observed signal.
Missed events and re-excitation allow later counts from other cycles.
Exchanging the channels can give a different function.
A normalized correlation is undefined if either steady-state current vanishes.
The difference between $\gtwo_{i\to s}$ and $\gtwo_{s\to i}$ expresses the time ordering of the cascade.

The intrinsic signed delay is the signal emission time minus the idler emission time.
The recorded timestamp difference also contains the timing error.
Subtracting the uncorrelated coincidence background gives the two-sided connected density,
\begin{align}
 C_{is}^{\pm}(\tau)=I_iI_s
 \begin{cases}
  \gtwo_{i\to s}(\tau)-1,&\tau\geq0,\\
  \gtwo_{s\to i}(-\tau)-1,&\tau<0.
 \end{cases}
\end{align}
Let the two timestamp errors be $\xi_i$ and $\xi_s$.
The recorded delay is $t=\tau+\xi_s-\xi_i$.
The response $h$ is the normalized probability density of $\xi_s-\xi_i$.
Convolution gives
\begin{align}
  H_{is}(t) = I_i I_s + \int_{-\infty}^{\infty} \dd{\tau} \, h(t-\tau) C_{is}^{\pm}(\tau).
  \label{eq:detector-histogram}
\end{align}
Here $t$ is the measured delay after the timing error.
The plotted normalized histogram is $\gtwo_{is,\mathrm{det}}(t)=H_{is}(t)/(I_iI_s)$.
The calculation uses a Gaussian two-channel delay response with full width at half maximum $\mathrm{\DetectorResponseFwhm}$.
This width describes the delay difference, not the timing uncertainty of each detector separately.
This assumed width is broader than the single-detector timing response demonstrated in Ref.~\cite{Korzh2020}.
That measurement does not give the joint response in the two cascade bands.
The symmetric coincidence window is $\mathcal{W}=[-T_{\mathrm{w}},T_{\mathrm{w}}]$, with halfwidth $T_{\mathrm{w}}=\mathrm{\CoincidenceWindowHalfWidth}$.
The accidental coincidence rate is the rate expected for uncorrelated idler and signal detections in this window,
\begin{align}
  R_{\mathrm{acc}} = 2T_{\mathrm{w}} I_i I_s.
  \label{eq:accidental-rate}
\end{align}
The connected coincidence rate measures the excess over this uncorrelated background,
\begin{align}
  R_{\mathrm{pair}} = \int_{\mathcal{W}} \dd{t}
  \left[H_{is}(t) -I_i I_s \right].
  \label{eq:coincidence-metrics}
\end{align}
It approaches the flux of emitted pairs when separate cascades dominate and the window captures their correlations.
It is not a branching fraction or a direct count of jointly emitted pairs at arbitrary drive.

The raw coincidence-to-accidental ratio compares the total coincidence rate with the accidental background,
\begin{align}
  \mathrm{CAR} = \frac{R_{\mathrm{pair}} + R_{\mathrm{acc}}}{R_{\mathrm{acc}}}.
  \label{eq:coincidence-to-accidental-ratio}
\end{align}
The conditional signal-count excess $\Delta N_{s|i}$ is the mean additional number of signal counts in the window per idler count, after subtraction of the uncorrelated background,
\begin{align}
  \Delta N_{s|i} = \frac{R_{\mathrm{pair}}}{I_i}.
  \label{eq:conditional-signal-excess}
\end{align}
It is a mean count, not the probability of at least one or of exactly one signal count, unless multiple counts and re-excitation within the window are negligible.
Independent external idler and signal efficiencies multiply the connected coincidence rate by their product.
External signal efficiency multiplies $\Delta N_{s|i}$ once.
The external efficiencies cancel from CAR if dark counts and detector dead time are negligible.

The spectrum of photon-count fluctuations in channels $a$ and $b$ is
\begin{align}
 S_{ab}(\w)=I_a\delta_{ab}
 +\int_{-\infty}^{\infty}\dd{\tau}\,e^{-i\w\tau}C_{ab}^{\pm}(\tau).
\end{align}
The term $I_a\delta_{ab}$ is the shot noise from individual detections and contributes only when $a=b$.
The connected correlation $C_{ab}^{\pm}$ describes correlations between distinct detections, with the uncorrelated background $I_aI_b$ subtracted.
Positive delay means that detection in channel $a$ precedes detection in channel $b$.
Here $C_{ab}^{\pm}$ extends the signed-delay definition above to any channel pair.
For the count record $j_a(u)=\sum_k\delta(u-t_{a,k})$, with mean $I_a$, the steady-state covariance is
\begin{align}
\begin{aligned}
 K_{ab}(\tau)&=\langle[j_a(u)-I_a][j_b(u+\tau)-I_b]\rangle\\
 &=I_a\delta_{ab}\delta(\tau)+C_{ab}^{\pm}(\tau).
\end{aligned}
\end{align}
The normally ordered correlation counts distinct events and therefore excludes the delta term.
The normalized spectrum is
\begin{align}
 \tilde{S}_{ab}(\w)=\frac{S_{ab}(\w)}{\sqrt{I_aI_b}}.
\end{align}
The row index is the earlier channel at positive delay.
This is the transpose of the later-first index convention in Sec.~VI.C of Ref.~\cite{Landi2024}.
The full matrix is complex and Hermitian, and $S_{ab}(-\w)=S_{ab}(\w)^*$.
The currents and spectra have units of inverse time, whereas $K_{ab}$ has units of inverse time squared.
A Poisson auto-spectrum equals its current, with no additional factor of two.
The angular frequency $\w$ is Fourier-conjugate to the delay $\tau$ and is plotted in energy units as $\hbar\w$.
This photon-count spectrum describes temporal fluctuations in the current, whereas an optical emission spectrum resolves emitted photon energies~\cite{Landi2024}.
The photon-count spectrum also differs from symmetrized quantum-current noise.
The single-channel Fano spectrum is
\begin{align}
 F_a(\w)=\frac{\operatorname{Re}S_{aa}(\w)}{I_a}.
 \label{eq:fano-spectrum}
\end{align}
Independent Poisson counts give unit Fano spectra.
The zero-frequency Fano factor is defined directly from the number $N_a(t_{\mathrm{obs}})$ of photons counted during an observation time $t_{\mathrm{obs}}$,
\begin{align}
  F_a(0) = \lim_{t_{\mathrm{obs}} \to \infty}
  \frac{\operatorname{Var}[N_a(t_{\mathrm{obs}})]}
  {\operatorname{E}[N_a(t_{\mathrm{obs}})]}.
  \label{eq:fano-zero}
\end{align}
The idler--signal cross-spectrum at zero frequency, $S_{is}(0)$, is the integral of $C_{is}^{\pm}$ over all delays, or equivalently the long-time growth rate of the count covariance~\cite{Landi2024}.
At zero frequency the imaginary part of $S_{ab}$ vanishes, so $\tilde{S}_{ab}(0)$ is real.
For independent external loss, the recorded photon current is
\begin{align}
 I_a^{\mathrm{rec}}=\zeta_a I_a,
 \label{eq:recorded-current}
\end{align}
where $\zeta_a$ is the probability, defined in Sec.~\ref{sec:counted-currents}, that an antenna photon in channel $a$ is recorded.
The same random loss changes the recorded count-noise spectrum,
\begin{align}
 S_{ab}^{\mathrm{rec}}=\zeta_a\zeta_b S_{ab}
 +\delta_{ab}\zeta_a(1-\zeta_a)I_a.
 \label{eq:recorded-noise}
\end{align}
The additional diagonal term accounts for fluctuations in whether each photon is recorded.
The normalized $g^{(2)}$ is unchanged by such loss.
Optical filtering acts on the emitted fields and can change the correlations between photons of different colors~\cite{delValle2012,delValle2016Erratum}.
It is distinct from independent collection loss and from jitter of timestamps already assigned to detected photons.
The Gaussian response used here multiplies the regular idler--signal cross-spectrum by $\exp[-\sigma_{\mathrm{diff}}^2\w^2/2]$, where $\sigma_{\mathrm{diff}}$ is the standard deviation of the timestamp difference.
An electronic filter instead smooths the count pulses themselves and also changes their shot-noise spectrum.
The optical and electronic responses of an actual detection system are not specified.

\subsection{Emitted-light spectrum}

The incoherent optical spectrum in detected channel $a$ is~\cite{Mollow1969,Carmichael1993}
\begin{align}
  S_{a,\mathrm{inc}}(\w_{\mathrm{opt}}) = 2\operatorname{Re} \int_0^{\infty} \dd{\tau} e^{-i\w_{\mathrm{opt}}\tau}
  \left\langle \Delta\hat{L}_a^{\dagger}(\tau) \, \Delta \hat{L}_a(0) \right\rangle .
  \label{eq:incoherent-spectrum}
\end{align}
Here $\Delta\hat{L}_a(t)=\hat{L}_a(t)-\langle\hat{L}_a(t)\rangle$ describes fluctuations about the mean field.
The subtracted mean gives the coherent part of the emission~\cite{Ficek2009}.
The operators are written in the laboratory frame, and $\w_{\mathrm{opt}}$ is the optical angular frequency.
This frequency labels the emitted field, whereas $\w$ in Sec.~\ref{sec:noise} labels fluctuations of the count record.

The plotted photon flux per unit photon energy $E$ is
\begin{align}
 \Phi_a(E)=\frac{S_{a,\mathrm{inc}}(E/\hbar)}{2\pi\hbar}.
 \label{eq:antenna-photon-flux}
\end{align}
For the steady-state Markov spectrum, the normalization is
\begin{align}
 \begin{aligned}
  \int_{-\infty}^{\infty}\dd{E}\,\Phi_a(E)
  &=\left\langle\Delta\hat{L}_a^{\dagger}\Delta\hat{L}_a\right\rangle_{\mathrm{ss}}\\
  &=I_a-\left|\langle\hat{L}_a\rangle_{\mathrm{ss}}\right|^2.
 \end{aligned}
 \label{eq:optical-spectrum-normalization}
\end{align}
Here $I_a$ is the total antenna-output photon current, and $|\langle\hat{L}_a\rangle_{\mathrm{ss}}|^2$ is its coherent contribution.
Their difference is the incoherent photon current.
Dividing this current by $I_a$ gives the incoherent fraction.
The coherent contribution is quoted as a fraction rather than as a continuous spectral density.

\begin{figure*}[tb]
\centering
\includegraphics{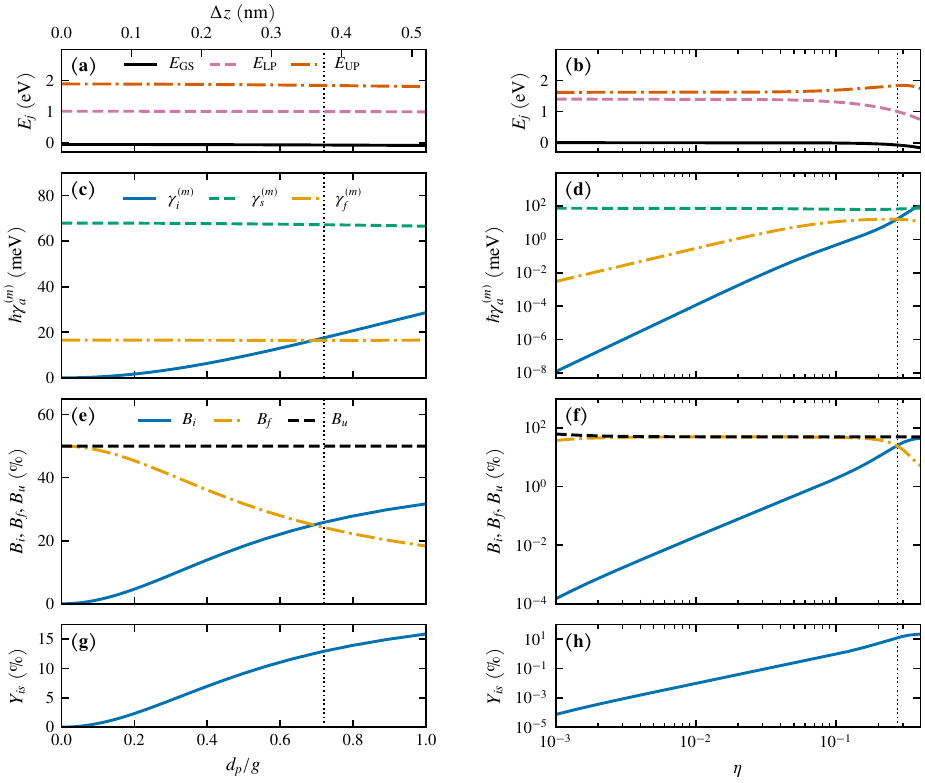}
\caption{
(a),(b) Undriven eigenenergies $E_j$ of GS, LP and UP, with the fixed energy zero of Hamiltonian~\eqref{eq:system-hamiltonian}.
(c),(d) Localized-mode loss rates $\hbar\gamma_a^{(m)}$ [Eq.~\eqref{eq:mode-rate}] for the idler ($a=i$, solid blue), signal ($a=s$, dashed green) and fluorescence ($a=f$, dash-dotted orange).
(e),(f) Antenna-output idler and fluorescence branching fractions $B_i$ and $B_f$, and unobserved upper-polariton decay probability $B_u$ [Eq.~\eqref{eq:branching-fractions}].
(g),(h) Pair yield $Y_{is}$ [Eq.~\eqref{eq:pair-yield}] per prepared UP, with the laser off.
The left column varies permanent-dipole asymmetry $d_p/g$ at the reference relative coupling $\eta=\RelativeCoupling$.
The top axis gives electron--hole separation $\Delta z$ for the fixed transition dipole $\mu_{eg}=\mathrm{\TransitionDipole}$ [Eq.~\eqref{eq:dipole-geometry}].
The right column varies $\eta$ at the reference displacement $\Delta z=\mathrm{\Displacement}$, with a logarithmic coupling axis and logarithmic vertical axes in (d),(f),(h).
Vertical dotted lines mark the values held fixed in the opposite column: $d_p/g=\PermanentDipoleRatio$ on the left and $\eta=\RelativeCoupling$ on the right.
All panels use fixed reference reservoir weights.
}\label{fig:currents}
\end{figure*}

To estimate the background from off-resonant laser excitation of LP, we retain the full cosine drive on both GS--LP and GS--UP transitions.
We set $\Delta z=0$, where the radiative idler transition is forbidden, and keep the loss operators obtained from the undriven dressed states.
After transients have decayed, we average the LP$\to$GS emission spectrum over one drive period.
An ideal signal filter transmits positive photon energies below the midpoint between the undriven signal energy and the pump energy.
Integrating the spectral photon flux over this band gives the background current, which we compare with the reference cascade signal current.

\section{Results}\label{sec:results}

The numerical calculations use QuTiP~\cite{Johansson2013,Lambert2026}.
We first diagonalize the undriven Hamiltonian in a basis of \PlasmonBasisSize{} plasmon number states.
We identify GS, LP and UP along the parameter scans as described in Sec.~\ref{sec:selection-rule}.
We retain these three dressed states in the driven dynamics.
Their energies and optical matrix elements, together with the reservoir spectra, give the transition rates entering the driven three-state master equation.
We use its analytical steady-state solution to calculate the photon currents and determine the state conditioned on a detection.
Propagating this conditional state gives the ordered correlations, while solving linear systems built from the same Liouvillian gives the count-noise spectra.

We use mode and exciton energies $\hbar\w_m=\mathrm{\ModeEnergy}$ and $\hbar\w_e=\mathrm{\ExcitonEnergy}$, relative coupling $\eta=\RelativeCoupling$, and localized-mode loss linewidth $\Gamma_m=\mathrm{\PlasmonLinewidth}$, following Ref.~\cite{Pompe2023}.
The reference calculation assumes a transition dipole $\mu_{eg}=\mathrm{\TransitionDipole}$, electron--hole separation $\Delta z=\mathrm{\Displacement}$ and antenna radiative fraction $r_m=\AntennaFraction$.
The reference resonant drive is $\hbar\Omega_{\mathrm{R}}=\mathrm{\ReferenceDriveEnergy}$.

The direct-emitter radiative linewidth $\Gamma_{\mathrm{E}}=\mathrm{\EmitterRadiativeLinewidth}$ follows from the assumed transition dipole in a homogeneous medium of reference refractive index $n_{\mathrm{ref}}=\ReferenceRefractiveIndex$.
The corresponding radiative lifetime is $\mathrm{\TransitionDipoleLifetime}$.
These estimates exclude nanogap enhancement, because the explicitly retained localized mode carries the enhanced coupling and the nanogap's effect on the remaining background has not been calculated.
We set the additional emitter nonradiative linewidth $\Gamma_{\mathrm{N}}$ to zero and omit residual population noise from the steady-state dynamics.

\subsection{Parity breaking and radiative photon currents}\label{sec:currents}

The permanent dipole opens the radiative idler transition (Fig.~\ref{fig:currents}).
At $d_p=0$ the steady-state model leaves the lower polariton empty, so the allowed signal transition carries no current and only upper-polariton fluorescence remains.
Connected idler--signal coincidences then vanish, and their normalized statistics are undefined.
In the dephasing stress test, population transfer from UP to LP grows with the dephasing strength, while the radiative idler count is zero at every strength, as parity requires.

At the reference separation, $d_p/g=\PermanentDipoleRatio$, and the calculated total decay rates give $\hbar\gamma_i=\mathrm{\ReferenceIdlerDecayEnergy}$, $\hbar\gamma_s=\mathrm{\ReferenceSignalDecayEnergy}$ and $\hbar\gamma_f=\mathrm{\ReferenceFluorescenceDecayEnergy}$.
The idler, signal and fluorescence wavelengths are $\mathrm{\IdlerWavelength}$, $\mathrm{\SignalWavelength}$ and $\mathrm{\FluorescenceWavelength}$, respectively.
The two cascade photons therefore occupy separate spectral bands.
The pump energy lies above the GaAs band gap and, for small Al fractions, above the AlGaAs gap~\cite{Vurgaftman2001}.
Absorption of the pump in the semiconductor is not included in the model.

At fixed reference reservoir weights, the localized-mode idler rate $\gamma_i^{(m)}$ of Eq.~\eqref{eq:mode-rate} and the antenna-output idler branching fraction $B_i$ of Eq.~\eqref{eq:branching-fractions} grow quadratically from zero at small displacement (Fig.~\ref{fig:currents}(c),(e)).
At larger displacement $\gamma_i^{(m)}$ falls below its quadratic extrapolation, and $B_i$ grows more slowly because the idler rate also enters the total upper-polariton loss rate in its denominator.
The permanent dipole changes all three rates through the dressed eigenstates.
Across the plotted displacement range, the localized-mode signal and fluorescence rates change only slightly (Fig.~\ref{fig:currents}(c)).
The increase in antenna idler branching is accompanied mainly by a decrease in antenna fluorescence branching (Fig.~\ref{fig:currents}(e)).

The reference calculation gives an antenna-output idler branching fraction of \IdlerBranchingPercent{}.
The antenna-output idler branching fraction is smaller than the mode branching fraction $B_m$ of Eq.~\eqref{eq:mode-branching} mainly because only the fraction $r_m$ of localized-mode loss leaves through the antenna.
This estimate depends on the unmeasured relative reservoir weights at the idler and fluorescence energies.
The idler and signal currents remain nearly equal, while the fluorescence current is of the same order.

The antenna-output currents correspond to $\mathrm{\IdlerRatePerSecond}$ for the idler and $\mathrm{\FluorescenceRatePerSecond}$ for fluorescence.
These are source currents, with the antenna radiative fraction included.
The external collection and detector efficiencies needed to predict measured count rates are unknown.

To compare exact parity with the reference permanent dipole, we keep $\Omega_{\mathrm{R}}/\gamma_f=\DriveRabiRatio{}$ fixed, with $\gamma_f$ recalculated at each displacement.
With this drive, the fluorescence current at the reference permanent dipole is \FluorescenceCurrentChange{} lower than at exact parity.
This large reduction does not come from a large change of the dressed states.
It comes from the newly opened idler channel: at the reference separation $\gamma_i$ exceeds $\gamma_f$, so the upper polariton empties more than twice as fast as at exact parity.
At weak drive the upper-polariton population scales as the square of the drive over the total upper-polariton loss rate, so this added loss, not the small changes of the other rates, lowers the fluorescence current.

Along the coupling scan (Fig.~\ref{fig:currents}(b),(d),(f)), all three localized-mode rates change as the dressed eigenstates evolve.
At $d_p=0$ the radiative idler remains absent throughout.
At the smaller coupling $\eta=\AttainableCoupling$, below the conventional ultrastrong threshold $\eta=\UltrastrongThreshold$, the model gives an antenna-output idler branching fraction of \AttainableCouplingBranching{}.
Parity breaking opens the idler without requiring ultrastrong coupling.

\subsection{Photon-pair yield and emitted-light spectrum}\label{sec:pair-output}

At fixed reference reservoir weights, the pair yield $Y_{is}$ of Eq.~\eqref{eq:pair-yield} vanishes at zero permanent dipole and grows quadratically at small displacement (Fig.~\ref{fig:currents}(g)).
At the reference displacement and antenna radiative fraction, the reference rates give $Y_{is}=\OpticalPairYieldPercent$.
The corresponding signal escape probability after an antenna idler is $B_{s|\mathrm{LP}}=\OpticalSignalEscapePercent$.
The pair yield therefore includes an additional escape loss compared with the idler branching fraction.
The yield increases with asymmetry and with coupling along the two scans (Fig.~\ref{fig:currents}(g),(h)).
Every point uses the idler, signal and fluorescence rates calculated at that displacement and coupling, with fixed reservoir weights.
Because the conditional signal escape probability remains nearly constant, the pair yield follows the corresponding idler branching fraction with an additional factor of approximately one half.

\begin{figure}[tb]
\centering
\includegraphics{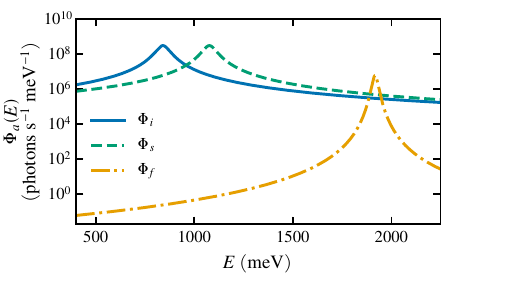}
\caption{Incoherent antenna emission flux $\Phi_a(E)$ [Eq.~\eqref{eq:antenna-photon-flux}] against photon energy $E$, under the reference continuous drive and with the reference rates.
The blue solid, green dashed and orange dash-dotted curves are idler, signal and fluorescence emission, respectively.
The elastic pump contribution is omitted.}
\label{fig:pair-output}
\end{figure}

\begin{figure*}[tb]
\centering
\includegraphics{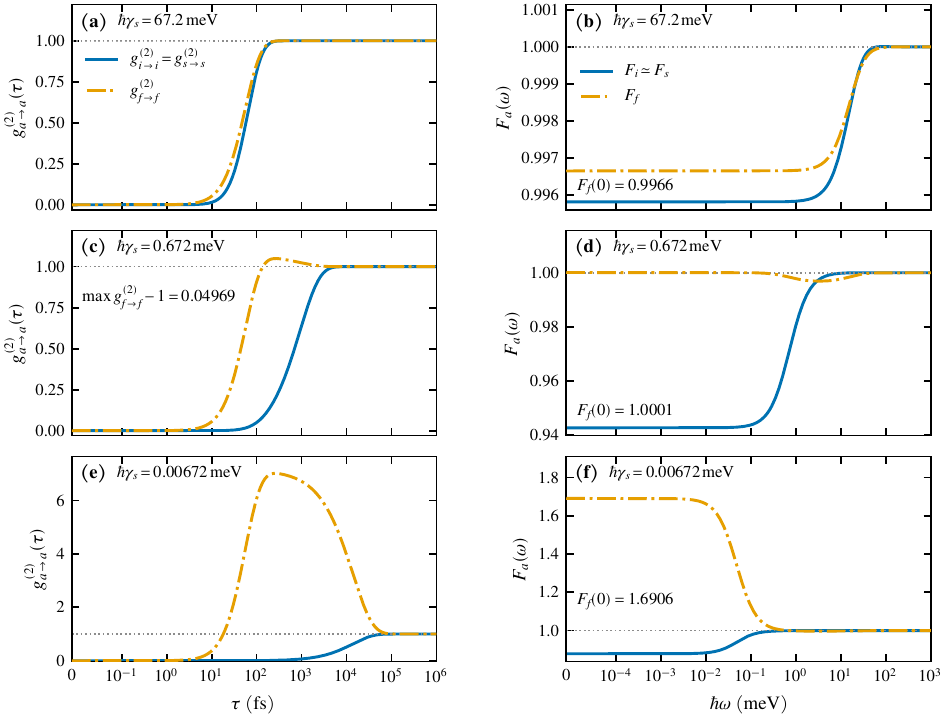}
\caption{Individual-channel statistics at drive $\hbar\Omega_{\mathrm{R}}=\mathrm{\ReferenceDriveEnergy}$ and displacement $\mathrm{\Displacement}$.
The rows use total signal decay rates $\hbar\gamma_s=\mathrm{\ReferenceSignalDecayEnergy}$, $\mathrm{\SlowSignalDecayEnergy}$ and $\mathrm{\IntermittentSignalDecayEnergy}$.
The idler and fluorescence rates retain the reference values given in Sec.~\ref{sec:currents}.
The signal-rate choices compare different assumed reservoir weights at the signal energy, with the Hamiltonian and counted fractions fixed.
(a),(c),(e) Normalized autocorrelations $\gtwo_{a\to a}(\tau)$ [Eq.~\eqref{eq:ordered-correlation}] versus intrinsic time delay $\tau$.
(b),(d),(f) Single-channel Fano spectra $F_a(\w)$ [Eq.~\eqref{eq:fano-spectrum}] versus $\hbar\w$, where $\w$ is the angular frequency of photon-count fluctuations.
Both horizontal axes are linear from zero to the first positive tick and logarithmic above it, allowing zero delay and zero frequency to be shown.
Grey dotted lines mark unity, the uncorrelated level for distinct-event correlations and the Poisson level for Fano noise.
Solid blue curves show the idler and dash-dotted orange curves show fluorescence.
Signal curves are not shown, because they coincide with the idler curves at this scale.
Annotations give the maximum of $\gtwo_{f\to f}(\tau)-1$ in (c) and the zero-frequency fluorescence Fano factors in (b),(d),(f).
All channels shown have nonzero currents.}
\label{fig:individual-statistics}
\end{figure*}

The emission flux $\Phi_a(E)$ in Fig.~\ref{fig:pair-output} separates the two cascade bands from fluorescence near the pump.
At the reference drive strength, \FluorescenceElasticFraction{} of the fluorescence is elastic scattering at the pump energy.
This coherent contribution is excluded from the plotted incoherent spectrum.
The idler and signal fluxes $\Phi_i(E)$ and $\Phi_s(E)$ contain no elastic component in the projected model.
Their integrated currents are nearly equal because the lower polariton is populated by the idler transition and depopulated by the signal transition.
Steady-state population balance equates these two total transition fluxes, although the signal rate is larger than the idler rate.
A spectrum alone cannot establish that photons in the two bands form pairs.
The ordered correlations below test their common cascade origin.

\subsection{Photon-count statistics}
\label{sec:statistics}

Individual-channel fluctuations show how excitation and decay set the intervals between repeated emissions.
Cross-correlations then test whether the two cascade channels occur in the expected order.

The idler and signal have identical normalized autocorrelations in every displayed case (Fig.~\ref{fig:individual-statistics}).
Independent photon loss leaves normalized autocorrelations unchanged.
If the idler and signal had equal counted fractions, their auto-Fano spectra would also be identical at every noise frequency $\w$.
The equality follows because each idler leaves the system in LP and each signal leaves it in GS, so both channels repeat the same cycle of waiting times, including coherent excitation and intervening fluorescence.
With the actual unequal counted fractions, the antenna-output Fano spectra differ slightly even though their normalized autocorrelations coincide.
That difference is unresolved at the scale of Fig.~\ref{fig:individual-statistics}.

\begin{figure}[!t]
\centering
\includegraphics{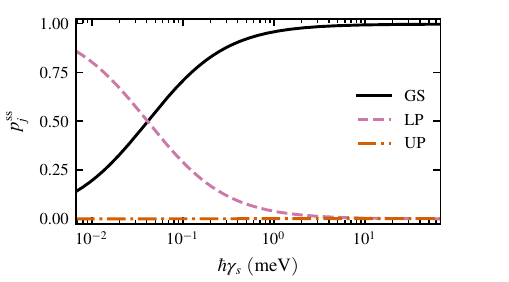}
\caption{
Steady-state occupations $p_j^{\mathrm{ss}}$ [Eq.~\eqref{eq:steady-state-occupation}] versus total signal decay rate $\hbar\gamma_s$.
The solid black, dashed purple and dash-dotted red curves show GS, LP and UP, respectively.
The logarithmic signal-rate axis spans the three cases in Fig.~\ref{fig:individual-statistics}.
The Hamiltonian, physical drive, idler and fluorescence rates, and counted fractions have their reference values.
}
\label{fig:steady-state-occupations}
\end{figure}

At the reference rates, the idler and fluorescence counts are close to Poissonian.
Over the three displayed signal rates, slower signal decay lowers the idler Fano factor and raises the fluorescence Fano factor.
At the bottom-row signal rate they reach $\IntermittentCascadeFano$ and $\IntermittentFluorescenceFano$, and the steady-state LP occupation is $\IntermittentLPOccupancy$.
Fluorescence occurs during the periods spent outside LP and is interrupted by occupation of LP.
The enhanced probability of another fluorescence photon after a finite delay is the delayed bunching seen in the autocorrelation.
This delayed bunching explains the larger variance of the fluorescence counts.
The cascade rate falls as LP occupation increases.
At the reference drive, slowing signal decay transfers steady-state occupation from GS to LP, while the UP occupation remains small throughout the range (Fig.~\ref{fig:steady-state-occupations}).

\begin{figure*}[!t]
\centering
\includegraphics{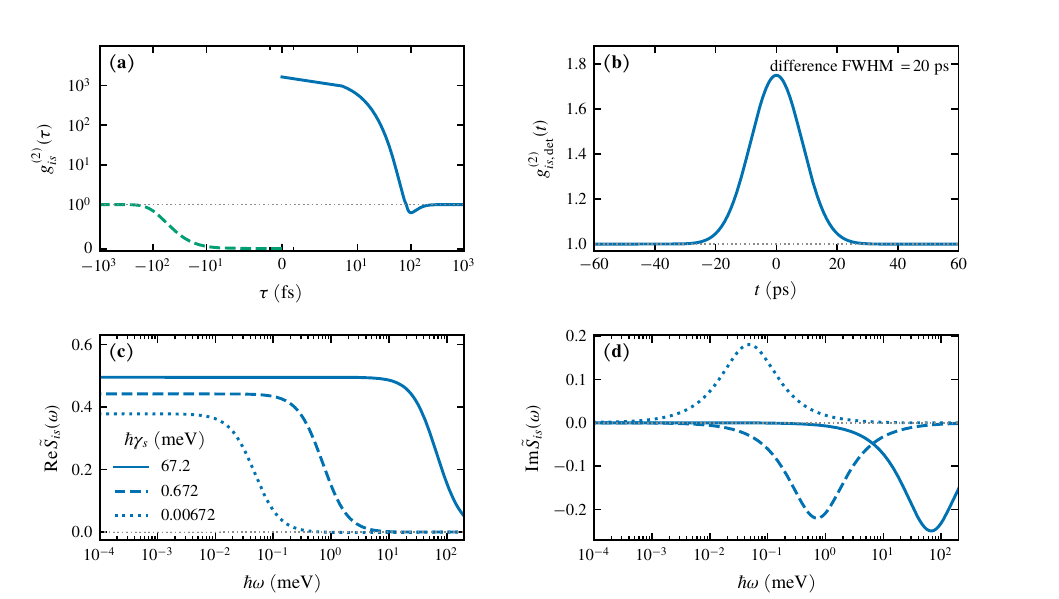}
\caption{Cascade order and the assumed timing response.
(a) Signed intrinsic correlation $\gtwo_{is}(\tau)$ at the reference parameters versus $\tau$, the signal emission time minus the idler emission time, equal to $\gtwo_{i\to s}(\tau)$ for $\tau\geq0$ and to $\gtwo_{s\to i}(-\tau)$ for $\tau<0$ [Eq.~\eqref{eq:ordered-correlation}].
Negative delays compare a signal photon from one cascade with an idler photon from a later cascade.
The two one-sided limits at zero are drawn separately.
In (a), the delay axis is linear within $\pm\mathrm{\OrderingLinearDelay}$ and logarithmic beyond, and the vertical axis is linear from zero to one and logarithmic above.
(b) Normalized coincidence histogram $\gtwo_{is,\mathrm{det}}(t)$ [Eq.~\eqref{eq:detector-histogram}] versus recorded delay $t$ at exactly the same source parameters, after a Gaussian timestamp-difference response of FWHM $\mathrm{\DetectorResponseFwhm}$.
The delay units change from fs in (a) to ps in (b).
(c),(d) Real and imaginary parts of the normalized idler--signal cross-spectrum $\tilde{S}_{is}(\w)=S_{is}(\w)/\sqrt{I_iI_s}$ for the three signal-rate cases of Fig.~\ref{fig:individual-statistics}, where $I_i$ and $I_s$ are the photon currents.
This normalization makes the spectrum dimensionless without fixing its peak or area to unity.
Independent channels give zero.
The horizontal axis is $\hbar\w$, where $\w$ is the angular frequency of photon-count fluctuations.
Solid, dashed and dotted lines identify the top-row, middle-row and bottom-row signal rates, respectively.
Horizontal grey dotted guides mark unity in (a),(b) and zero in (c),(d).
The intrinsic spectra are shown before timestamp jitter.}
\label{fig:statistics}
\end{figure*}

An idler transition leaves the system in LP, whereas a signal transition leaves it in GS.
The idler-then-signal correlation is therefore large at zero delay and returns to one within the LP lifetime, $\mathrm{\LPLifetime}$, while the signal-then-idler correlation vanishes at zero delay (Fig.~\ref{fig:statistics}(a)).

The assumed timestamp response spreads the zero-delay peak of the idler-then-signal correlation into a much lower and broader maximum of the recorded histogram (Fig.~\ref{fig:statistics}(b)).
The response leaves the integrated connected area unchanged but does not resolve the intrinsic emission order.
Because the response is far wider than the LP lifetime $\mathrm{\LPLifetime}$, the recorded histogram is the response function scaled by the connected area, and its shape carries no information about the source.
With dark counts and dead time neglected, the raw CAR is \CoincidenceToAccidentalRatio{}.
The conditional signal-count excess is $\Delta N_{s|i}=\ConditionalSignalExcess{}$ signal counts per antenna idler.
External signal loss reduces $\Delta N_{s|i}$.
Reducing the drive lowers brightness while increasing CAR, because the accidental rate is quadratic in the currents whereas the connected coincidence rate is linear in them.

The normalized idler--signal cross-spectrum is the cross-channel counterpart of the single-channel Fano spectrum.
At the reference drive it is positive at zero frequency (Fig.~\ref{fig:statistics}(c)).
A positive zero-frequency cross-spectrum alone establishes neither cascade order nor entanglement.
The part of the connected delay correlation that is symmetric under $\tau\to-\tau$ gives the real part of the cross-spectrum, while the antisymmetric part gives its imaginary part (Fig.~\ref{fig:statistics}(c),(d)).
A finite imaginary part therefore reflects unequal correlations at positive and negative delays.
The imaginary part vanishes at zero frequency.
Re-excitation and missed photons permit correlations between different cycles, so a sign change of the imaginary part does not mean that the transitions within one cascade occur in reverse order.
The assumed timestamp response suppresses features at frequencies above its characteristic response scale.
A device-specific two-band timing response is required to predict their visibility.

\begin{figure*}[tb]
\centering
\includegraphics{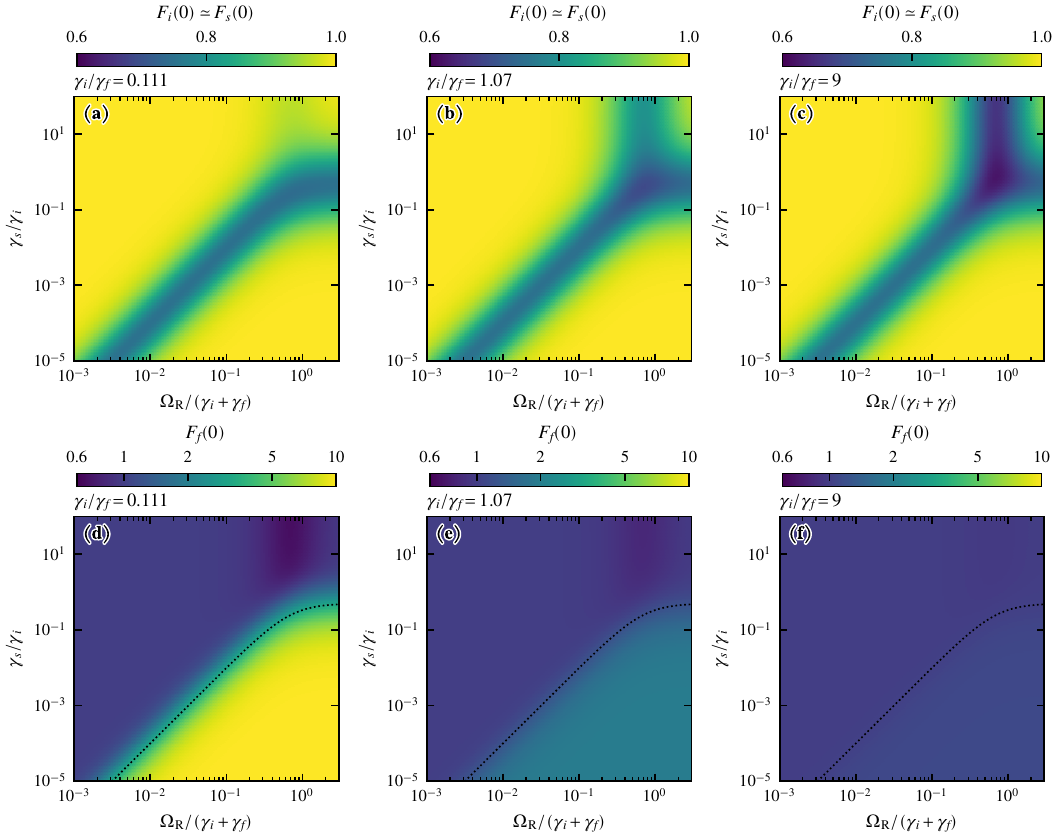}
\caption{Drive, signal decay and branching competition in the effective model.
(a)--(c) Zero-frequency Fano factor $F_i(0)$ [Eq.~\eqref{eq:fano-zero}] of the antenna-output idler counts.
(d)--(f) Zero-frequency Fano factor $F_f(0)$ of the antenna-output fluorescence counts.
Columns compare different assumed reservoir weights at the idler and fluorescence energies.
The total idler-to-fluorescence rate ratios are labeled, with reference branching in the center.
All panels use drive divided by total UP loss rate horizontally and the signal-to-idler decay-rate ratio vertically.
The total UP loss rate is fixed at $\hbar\gamma_{\mathrm{UP}}=\mathrm{\BranchMapEscapeEnergy}$.
The Hamiltonian, displacement $\mathrm{\Displacement}$ and counted fractions remain fixed.
Each observable has a common color scale across columns, linear for cascade noise and logarithmic for fluorescence noise.
Dotted curves mark a steady-state LP occupation of one half, with higher occupation below them, and coincide across columns in these coordinates.
}
\label{fig:rate-maps}
\end{figure*}

At the fixed reference asymmetry, Fig.~\ref{fig:rate-maps} compares assumed reservoir weights through three dimensionless ratios.
The drive over the total UP loss rate compares excitation of UP with its decay.
The signal over the idler rate sets how long LP stores the excitation, and the idler over the fluorescence rate sets the branching of UP decay.
At fixed coordinates, the steady-state LP occupation is independent of the branching ratio, whereas both Fano factors change.
Counting every transition, including unobserved loss, gives exactly equal idler and signal Fano factors.
The antenna-output idler and signal Fano factors differ only through the slightly unequal counted fractions, so Fig.~\ref{fig:rate-maps}(a)--(c) also describes the signal channel.
Stronger preference for idler decay lowers the minimum cascade Fano factor and reduces the fluorescence excess noise in the slow-signal limit.
At reference branching, the minimum over drive and signal rate is $\BranchMapReferenceMinimum$, at drive $\hbar\Omega_{\mathrm{R}}=\mathrm{\BranchMapOptimalDriveEnergy}$.
At the lower reference drive of Fig.~\ref{fig:individual-statistics}, the idler Fano factor is smallest at an intermediate signal rate.
Slower signal decay beyond that minimum makes the signal waiting stage dominate the cycle duration and moves the cascade noise back toward its Poisson limit.
A Fano factor measures fluctuations relative to the mean count, not brightness.
The LP-occupancy contours show where slow signal decay stores the excitation in LP and the currents are low.
With the total UP loss rate fixed as in Fig.~\ref{fig:rate-maps}, the highest signal rates broaden the idler and signal lines until the two bands are no longer separated.
The omitted off-resonant GS--LP drive can substantially change the photon currents at the extreme slow-signal corner, where slow signal decay stores the excitation in LP and the off-resonant laser coupling acts on the accumulated population.
These current comparisons do not bound errors in the noise or establish the validity of the whole plotted domain.

\subsection{Background from the omitted drive term}
\label{sec:offresonant}

The projected drive omits an off-resonant ground-to-lower-polariton term whose matrix element is larger than the retained resonant one.
A periodic three-state calculation at $\Delta z=0$ shows that almost all of the emitted light is elastic scattering at the pump.
A signal-band filter leaves a background of $\OffResonantFilteredPercent$ of the cascade signal current.
Spectral filtering is therefore part of the measurement definition.
The periodic calculation tests background from the retained radiative transitions at the stated drive.
It does not exclude every possible emission process in the idler band.
The period-averaged lower-polariton population agrees with its leading-order weak-drive estimate.
This calculation does not give the photon-count correlations behind finite-bandwidth filters in the idler and signal bands.

\section{Discussion}
\label{sec:discussion}

\subsection{Operating principle and relation to prior work}
\label{sec:interpretation}

The proposed source uses the permanent exciton dipole to make the upper-to-lower polariton transition radiative.
The transition then prepares the lower polariton for signal emission.
The ordered correlation identifies this sequence within the effective model (Fig.~\ref{fig:statistics}(a)), while the dephasing stress test separates photon emission from population transfer.
The idler rate depends on the coupling and on the optical reservoirs.
The pair yield includes escape loss at both steps, so an idler branching fraction alone overestimates the probability of two antenna photons.
Because $Y_{is}$ is proportional to $r_m^2$ and depends on the unmeasured reservoir weights (Sec.~\ref{sec:currents}), the unknown antenna response is a main uncertainty in the output.
The separated emission bands identify the two channels (Fig.~\ref{fig:pair-output}).

The device builds on the gap-plasmon proposal of Pompe \textit{et al.}~\cite{Pompe2023} and the permanent-dipole emission mechanism of Scala \textit{et al.}~\cite{Scala2021}.
We study this cascade in an effective model of a semiconductor quantum dot and calculate the pair yield and photon-count correlations.

\subsection{Predicted detection signatures}
\label{sec:detection-signatures}

The detection model predicts an excess of idler-signal coincidences in separate spectral bands.
The assumed response broadens this excess into an unresolved peak, and weaker driving improves CAR at the cost of connected coincidence rate.
The quoted CAR and conditional signal-count excess neglect dark counts and dead time.
Collection efficiencies and numerical aperture are unspecified, so neither detector count rates nor the integration time needed to resolve the excess are predicted.

\subsection{Feasibility and limitations}
\label{sec:limitations}

Room-temperature plasmonic strong coupling has been measured~\cite{Chikkaraddy2016,Gross2018,Hu2024}, while the cited light-hole transitions and permanent dipoles come from cryogenic structures~\cite{Huo2014,Huang2021}.
No cited device combines these ingredients with the proposed cascade.
The model has no temperature-dependent emitter rates.
The photons share one polarization, so polarization entanglement is not predicted.
The separated emission bands instead make the cascade a heralded pair source: detection of an idler photon predicts a signal photon in the other band, with the conditional signal-count excess quantified in Sec.~\ref{sec:statistics}.

Displacement, transition dipole, coupling strength and antenna radiative fraction are assumed independently.
Absorption of the retained mode is included.
The antenna spectrum, dot placement and device variability remain uncharacterized.
Frequency wandering (spectral diffusion), intermittent emission (blinking), heating and absorption of the pump in the semiconductor are omitted.
The projected Rabi frequency alone does not specify incident power.
The broad rate scan omits off-resonant GS--LP driving and transitions outside the retained states.
Stronger driving can populate omitted states and introduce off-resonant background.
The retained transition separations do not change across the map, but the optical linewidths, and the finite response time of real filters, still limit how well individual photons can be assigned to transitions.
The ideal spectral-edge check at the reference drive does not validate finite optical filters across the rate map.
A prediction for those cases requires device reservoir weights, additional loss and noise rates, and a field-level filter calculation.
The map therefore establishes properties of the effective model only, not a validated device operating range or independent device controls.

The reference coupling $\eta=\RelativeCoupling$ lies near the upper end of the perturbative ultrastrong range, $g/\w_m=\UltrastrongThreshold$--$\UltrastrongPerturbativeLimit$ at resonance~\cite{FornDiaz2019}.
The corresponding LP--UP splitting is \SplittingOverReported{} times the mean splitting $\mathrm{\DotGapSplitting}$ measured in scattering spectra of single colloidal dots~\cite{Hu2024}.
Larger relative couplings have been measured for collective nanorod modes in a Fabry-P\'{e}rot cavity~\cite{Baranov2020}.

With $\mathrm{\TransitionDipole}$, the reference coupling requires an estimated mode volume $\mathrm{\ImpliedModeVolume}$ (Sec.~\ref{sec:dressed-elements}).
A GaAs dot of the selected type is much larger than this volume.
Configuration-interaction models of such dots use a cone with a base diameter of $\mathrm{\DotBaseDiameter}$~\cite{Huang2021}, and the light-hole dots of Ref.~\cite{Huo2014} are about $\mathrm{\DotHeight}$ high.
The mode field must cover the dot, and a mode that fills a cylinder of this diameter and height gives $\eta=\DotCoupling$.
At the $\mathrm{\PublishedModeVolume}$ nanostar-gap volume of Pompe \textit{et al.}~\cite{Pompe2023}, the same dipole gives $\eta=\PublishedModeVolumeCoupling$.

\section{Conclusions}
\label{sec:conclusions}

A permanent exciton dipole enables cascaded photon-pair emission in the effective model of a GaAs quantum dot in a plasmonic nanogap.
At fixed reservoir weights, the idler rate and pair yield grow quadratically with electron--hole separation at small separation.
The mechanism does not require ultrastrong coupling.

Parity breaking supplies the radiative pathway, while competition with fluorescence and escape losses at both steps set the pair yield.
Slower signal decay can make the cascade counts more regular (sub-Poissonian) while increasing fluorescence fluctuations through longer occupation of LP.
The ordered correlation connects the two emission bands to the cascade, but the assumed detector timing leaves only an unresolved coincidence excess.
Reducing the drive improves coincidence contrast at the cost of connected coincidence rate.

The results establish an operating principle within the effective model.
The reference coupling exceeds the estimate for a GaAs dot of the reported size by a factor of \ReferenceOverDotCoupling{}.
The growth-axis transition dipole, additional emitter losses, antenna response and collection efficiencies remain unmeasured for the proposed device.
Device feasibility and coincidence measurability, including at room temperature, are not established.

\begin{acknowledgments}
This work was supported by the Marie Sk\l{}odowska-Curie Actions COFUND project, cofunded by the European Union (Physics for Future, Grant Agreement No. 101081515).
The author used AI tools (OpenAI GPT-5 and GPT-6, Anthropic Claude Opus 5 and Claude Sonnet 5, and GLM-5.3 hosted by Mistral) for assistance with code development and manuscript editing.
The author directed each coding and editing task and checked all output: the code against analytic limits and independent numerical checks, and the text statement by statement against its sources.
The author takes responsibility for the scientific content and final manuscript.
\end{acknowledgments}

\section*{Data availability}

The data necessary to reproduce the results of this article are available on Zenodo~\cite{reproduction-package}.

\bibliography{refs}

@article{Pompe2023,
  title = {Pure dephasing induced single-photon parametric down-conversion in a strongly coupled plasmon-exciton system},
  author = {Pompe, Ruben and Hensen, Matthias and Otten, Matthew and Gray, Stephen K. and Pfeiffer, Walter},
  journal = {Phys. Rev. B},
  volume = {108},
  pages = {115432},
  year = {2023},
  doi = {10.1103/PhysRevB.108.115432},
}

@article{Strauch1990,
  title = {Phonon dispersion in {GaAs}},
  author = {Strauch, D. and Dorner, B.},
  journal = {J. Phys.: Condens. Matter},
  volume = {2},
  pages = {1457--1474},
  year = {1990},
  doi = {10.1088/0953-8984/2/6/006},
}

@article{Landi2024,
  title = {Current fluctuations in open quantum systems: Bridging the gap between quantum continuous measurements and full counting statistics},
  author = {Landi, Gabriel T. and Kewming, Michael J. and Mitchison, Mark T. and Potts, Patrick P.},
  journal = {PRX Quantum},
  volume = {5},
  pages = {020201},
  year = {2024},
  doi = {10.1103/PRXQuantum.5.020201},
}

@article{Farina2019,
  title = {Open-quantum-system dynamics: Recovering positivity of the {R}edfield equation via the partial secular approximation},
  author = {Farina, Donato and Giovannetti, Vittorio},
  journal = {Phys. Rev. A},
  volume = {100},
  pages = {012107},
  year = {2019},
  doi = {10.1103/PhysRevA.100.012107},
}

@article{Trushechkin2021,
  title = {Unified {Gorini--Kossakowski--Lindblad--Sudarshan} quantum master equation beyond the secular approximation},
  author = {Trushechkin, Anton},
  journal = {Phys. Rev. A},
  volume = {103},
  pages = {062226},
  year = {2021},
  doi = {10.1103/PhysRevA.103.062226},
}

@article{Hopfield1958,
  title = {Theory of the contribution of excitons to the complex dielectric constant of crystals},
  author = {Hopfield, J. J.},
  journal = {Phys. Rev.},
  volume = {112},
  pages = {1555--1567},
  year = {1958},
  doi = {10.1103/PhysRev.112.1555},
}

@article{Ficek2009,
  title = {Comment on ``{D}iscrepancies in the resonance-fluorescence spectrum calculated with two methods''},
  author = {Ficek, Zbigniew},
  journal = {Phys. Rev. A},
  volume = {79},
  pages = {057401},
  year = {2009},
  doi = {10.1103/PhysRevA.79.057401},
}

@article{Mollow1969,
  title = {Power spectrum of light scattered by two-level systems},
  author = {Mollow, B. R.},
  journal = {Phys. Rev.},
  volume = {188},
  pages = {1969--1975},
  year = {1969},
  doi = {10.1103/PhysRev.188.1969},
}

@book{Carmichael1993,
  title = {An Open Systems Approach to Quantum Optics},
  author = {Carmichael, Howard J.},
  series = {Lecture Notes in Physics Monographs},
  volume = {18},
  publisher = {Springer Berlin, Heidelberg},
  year = {1993},
  doi = {10.1007/978-3-540-47620-7},
}

@article{Flindt2008,
  title = {Counting statistics of non-{Markovian} quantum stochastic processes},
  author = {Flindt, Christian and Novotn{\'y}, Tom{\'a}{\v{s}} and Braggio, Alessandro and Sassetti, Maura and Jauho, Antti-Pekka},
  journal = {Phys. Rev. Lett.},
  volume = {100},
  pages = {150601},
  year = {2008},
  doi = {10.1103/PhysRevLett.100.150601},
}

@article{Marcos2010,
  title = {Finite-frequency counting statistics of electron transport: {Markovian} theory},
  author = {Marcos, D. and Emary, C. and Brandes, T. and Aguado, R.},
  journal = {New J. Phys.},
  volume = {12},
  pages = {123009},
  year = {2010},
  doi = {10.1088/1367-2630/12/12/123009},
}

@article{Emary2007,
  title = {Frequency-dependent counting statistics in interacting nanoscale conductors},
  author = {Emary, C. and Marcos, D. and Aguado, R. and Brandes, T.},
  journal = {Phys. Rev. B},
  volume = {76},
  pages = {161404},
  year = {2007},
  doi = {10.1103/PhysRevB.76.161404},
}

@article{Johansson2013,
  title = {{QuTiP} 2: A {Python} framework for the dynamics of open quantum systems},
  author = {Johansson, J. R. and Nation, P. D. and Nori, Franco},
  journal = {Comput. Phys. Commun.},
  volume = {184},
  pages = {1234--1240},
  year = {2013},
  doi = {10.1016/j.cpc.2012.11.019},
}

@article{Lambert2026,
  title = {{QuTiP} 5: The Quantum Toolbox in {Python}},
  author = {Lambert, Neill and Gigu{\`e}re, Eric and Menczel, Paul and Li, Boxi and Hopf, Patrick and Su{\'a}rez, Gerardo and Gali, Marc and Lishman, Jake and Gadhvi, Rushiraj and Agarwal, Rochisha and Galicia, Asier and Shammah, Nathan and Nation, Paul and Johansson, J. R. and Ahmed, Shahnawaz and Cross, Simon and Pitchford, Alexander and Nori, Franco},
  journal = {Phys. Rep.},
  volume = {1153},
  pages = {1--62},
  year = {2026},
  doi = {10.1016/j.physrep.2025.10.001},
}

@article{Chikkaraddy2016,
  title = {Single-molecule strong coupling at room temperature in plasmonic nanocavities},
  author = {Chikkaraddy, Rohit and de Nijs, Bart and Benz, Felix and Barrow, Steven J. and Scherman, Oren A. and Rosta, Edina and Demetriadou, Angela and Fox, Peter and Hess, Ortwin and Baumberg, Jeremy J.},
  journal = {Nature},
  volume = {535},
  pages = {127--130},
  year = {2016},
  doi = {10.1038/nature17974},
}

@article{Lindblad1976,
  title = {On the generators of quantum dynamical semigroups},
  author = {Lindblad, G.},
  journal = {Commun. Math. Phys.},
  volume = {48},
  number = {2},
  pages = {119--130},
  year = {1976},
  doi = {10.1007/BF01608499},
}

@article{Braak2011,
  title = {Integrability of the {R}abi model},
  author = {Braak, D.},
  journal = {Phys. Rev. Lett.},
  volume = {107},
  pages = {100401},
  year = {2011},
  doi = {10.1103/PhysRevLett.107.100401},
}

@article{Settineri2018,
  title = {Dissipation and thermal noise in hybrid quantum systems in the ultrastrong-coupling regime},
  author = {Settineri, A. and Macr{\`\i}, V. and Ridolfo, A. and Di Stefano, O. and Frisk Kockum, A. and Nori, F. and Savasta, S.},
  journal = {Phys. Rev. A},
  volume = {98},
  pages = {053834},
  year = {2018},
  doi = {10.1103/PhysRevA.98.053834},
}

@article{Beaudoin2011,
  title = {Dissipation and ultrastrong coupling in circuit {QED}},
  author = {Beaudoin, F. and Gambetta, J. M. and Blais, A.},
  journal = {Phys. Rev. A},
  volume = {84},
  pages = {043832},
  year = {2011},
  doi = {10.1103/PhysRevA.84.043832},
}

@article{ChiSquared2023,
  title = {Degenerate parametric down-conversion facilitated by exciton--plasmon polariton states in a nonlinear plasmonic cavity},
  author = {Piryatinski, Andrei and Sukharev, Maxim},
  journal = {Nanotechnology},
  volume = {34},
  pages = {175001},
  year = {2023},
  doi = {10.1088/1361-6528/acb5a8},
}

@article{Rabi1936,
  title = {On the process of space quantization},
  author = {Rabi, I. I.},
  journal = {Phys. Rev.},
  volume = {49},
  pages = {324--328},
  year = {1936},
  doi = {10.1103/PhysRev.49.324},
}

@article{Xie2017,
  title = {The quantum {R}abi model: solution and dynamics},
  author = {Xie, Qiongtao and Zhong, Honghua and Batchelor, Murray T. and Lee, Chaohong},
  journal = {J. Phys. A: Math. Theor.},
  volume = {50},
  pages = {113001},
  year = {2017},
  doi = {10.1088/1751-8121/aa5a65},
}

@article{Benson2000,
  title = {Regulated and entangled photons from a single quantum dot},
  author = {Benson, Oliver and Santori, Charles and Pelton, Matthew and Yamamoto, Yoshihisa},
  journal = {Phys. Rev. Lett.},
  volume = {84},
  pages = {2513--2516},
  year = {2000},
  doi = {10.1103/PhysRevLett.84.2513},
}

@article{Burnham1970,
  title = {Observation of simultaneity in parametric production of optical photon pairs},
  author = {Burnham, David C. and Weinberg, Donald L.},
  journal = {Phys. Rev. Lett.},
  volume = {25},
  pages = {84--87},
  year = {1970},
  doi = {10.1103/PhysRevLett.25.84},
}

@article{Korzh2020,
  title = {Demonstration of sub-3 ps temporal resolution with a superconducting nanowire single-photon detector},
  author = {Korzh, Boris and Zhao, Qing-Yuan and Allmaras, Jason P. and Frasca, Simone and Autry, Travis M. and Bersin, Eric A. and Beyer, Andrew D. and Briggs, Ryan M. and Bumble, Bruce and Colangelo, Marco and Crouch, Garrison M. and Dane, Andrew E. and Gerrits, Thomas and Lita, Adriana E. and Marsili, Francesco and Moody, Galan and Pe{\~{n}}a, Cristi{\'a}n and Ramirez, Edward and Rezac, Jake D. and Sinclair, Neil and Stevens, Martin J. and Velasco, Angel E. and Verma, Varun B. and Wollman, Emma E. and Xie, Si and Zhu, Di and Hale, Paul D. and Spiropulu, Maria and Silverman, Kevin L. and Mirin, Richard P. and Nam, Sae Woo and Kozorezov, Alexander G. and Shaw, Matthew D. and Berggren, Karl K.},
  journal = {Nat. Photonics},
  volume = {14},
  pages = {250--255},
  year = {2020},
  doi = {10.1038/s41566-020-0589-x},
}

@article{Ekert1991,
  title = {Quantum cryptography based on {B}ell's theorem},
  author = {Ekert, Artur K.},
  journal = {Phys. Rev. Lett.},
  volume = {67},
  pages = {661--663},
  year = {1991},
  doi = {10.1103/PhysRevLett.67.661},
}

@article{Yin2020,
  title = {Entanglement-based secure quantum cryptography over 1,120 kilometres},
  author = {Yin, Juan and Li, Yu-Huai and Liao, Sheng-Kai and Yang, Meng and Cao, Yuan and Zhang, Liang and Ren, Ji-Gang and Cai, Wen-Qi and Liu, Wei-Yue and Li, Shuang-Lin and Shu, Rong and Huang, Yong-Mei and Deng, Lei and Li, Li and Zhang, Qiang and Liu, Nai-Le and Chen, Yu-Ao and Lu, Chao-Yang and Wang, Xiang-Bin and Xu, Feihu and Wang, Jian-Yu and Peng, Cheng-Zhi and Ekert, Artur K. and Pan, Jian-Wei},
  journal = {Nature},
  volume = {582},
  pages = {501--505},
  year = {2020},
  doi = {10.1038/s41586-020-2401-y},
}

@article{Xu2020,
  title = {Secure quantum key distribution with realistic devices},
  author = {Xu, Feihu and Ma, Xiongfeng and Zhang, Qiang and Lo, Hoi-Kwong and Pan, Jian-Wei},
  journal = {Rev. Mod. Phys.},
  volume = {92},
  pages = {025002},
  year = {2020},
  doi = {10.1103/RevModPhys.92.025002},
}

@article{Freedman1972,
  title = {Experimental test of local hidden-variable theories},
  author = {Freedman, Stuart J. and Clauser, John F.},
  journal = {Phys. Rev. Lett.},
  volume = {28},
  pages = {938--941},
  year = {1972},
  doi = {10.1103/PhysRevLett.28.938},
}

@article{Kwiat1995,
  title = {New high-intensity source of polarization-entangled photon pairs},
  author = {Kwiat, Paul G. and Mattle, Klaus and Weinfurter, Harald and Zeilinger, Anton and Sergienko, Alexander V. and Shih, Yanhua},
  journal = {Phys. Rev. Lett.},
  volume = {75},
  pages = {4337--4341},
  year = {1995},
  doi = {10.1103/PhysRevLett.75.4337},
}

@article{Akopian2006,
  title = {Entangled photon pairs from semiconductor quantum dots},
  author = {Akopian, N. and Lindner, N. H. and Poem, E. and Berlatzky, Y. and Avron, J. and Gershoni, D. and Gerardot, B. D. and Petroff, P. M.},
  journal = {Phys. Rev. Lett.},
  volume = {96},
  pages = {130501},
  year = {2006},
  doi = {10.1103/PhysRevLett.96.130501},
}

@article{Liu2019,
  title = {A solid-state source of strongly entangled photon pairs with high brightness and indistinguishability},
  author = {Liu, Jin and Su, Rongbin and Wei, Yuming and Yao, Beimeng and Covre da Silva, Saimon Filipe and Yu, Ying and Iles-Smith, Jake and Srinivasan, Kartik and Rastelli, Armando and Li, Juntao and Wang, Xuehua},
  journal = {Nat. Nanotechnol.},
  volume = {14},
  pages = {586--593},
  year = {2019},
  doi = {10.1038/s41565-019-0435-9},
}

@article{Wang2019,
  title = {On-demand semiconductor source of entangled photons which simultaneously has high fidelity, efficiency, and indistinguishability},
  author = {Wang, Hui and Hu, Hai and Chung, T.-H. and Qin, Jian and Yang, Xiaoxia and Li, J.-P. and Liu, R.-Z. and Zhong, H.-S. and He, Y.-M. and Ding, Xing and Deng, Y.-H. and Dai, Qing and Huo, Y.-H. and H{\"o}fling, Sven and Lu, Chao-Yang and Pan, Jian-Wei},
  journal = {Phys. Rev. Lett.},
  volume = {122},
  pages = {113602},
  year = {2019},
  doi = {10.1103/PhysRevLett.122.113602},
}

@article{Zhao2020,
  title = {High quality entangled photon pair generation in periodically poled thin-film lithium niobate waveguides},
  author = {Zhao, Jie and Ma, Chaoxuan and R{\"u}sing, Michael and Mookherjea, Shayan},
  journal = {Phys. Rev. Lett.},
  volume = {124},
  pages = {163603},
  year = {2020},
  doi = {10.1103/PhysRevLett.124.163603},
}

@article{Wu2026,
  title = {{P}urcell-enhanced two-photon emission from a quantum dot via dark-state biexciton loading},
  author = {Wu, Bang and Liu, Li and Liu, Hanqing and Mao, Xinrui and Wang, Xu-Jie and Ni, Haiqiao and Niu, Zhichuan and Yuan, Zhiliang},
  journal = {Nat. Mater.},
  volume = {25},
  pages = {595--601},
  year = {2026},
  doi = {10.1038/s41563-026-02522-9},
}

@article{JaynesCummings1963,
  title = {Comparison of quantum and semiclassical radiation theories with application to the beam maser},
  author = {Jaynes, E. T. and Cummings, F. W.},
  journal = {Proc. IEEE},
  volume = {51},
  pages = {89--109},
  year = {1963},
  doi = {10.1109/PROC.1963.1664},
}

@article{Baranov2020,
  title = {Ultrastrong coupling between nanoparticle plasmons and cavity photons at ambient conditions},
  author = {Baranov, Denis G. and Munkhbat, Battulga and Zhukova, Elena and Bisht, Ankit and Canales, Adriana and Rousseaux, Benjamin and Johansson, G{\"o}ran and Antosiewicz, Tomasz J. and Shegai, Timur},
  journal = {Nat. Commun.},
  volume = {11},
  pages = {2715},
  year = {2020},
  doi = {10.1038/s41467-020-16524-x},
}

@article{FornDiaz2019,
  title = {Ultrastrong coupling regimes of light-matter interaction},
  author = {Forn-D{\'\i}az, P. and Lamata, L. and Rico, E. and Kono, J. and Solano, E.},
  journal = {Rev. Mod. Phys.},
  volume = {91},
  pages = {025005},
  year = {2019},
  doi = {10.1103/RevModPhys.91.025005},
}

@article{DiStefano2019,
  title = {Resolution of gauge ambiguities in ultrastrong-coupling cavity quantum electrodynamics},
  author = {Di Stefano, Omar and Settineri, Alessio and Macr{\`\i}, Vincenzo and Garziano, Luigi and Stassi, Roberto and Savasta, Salvatore and Nori, Franco},
  journal = {Nat. Phys.},
  volume = {15},
  pages = {803--808},
  year = {2019},
  doi = {10.1038/s41567-019-0534-4},
}

@article{Savasta2021,
  title = {Gauge principle and gauge invariance in two-level systems},
  author = {Savasta, Salvatore and Di Stefano, Omar and Settineri, Alessio and Zueco, David and Hughes, Stephen and Nori, Franco},
  journal = {Phys. Rev. A},
  volume = {103},
  pages = {053703},
  year = {2021},
  doi = {10.1103/PhysRevA.103.053703},
}

@article{Dung1998,
  title = {Three-dimensional quantization of the electromagnetic field in dispersive and absorbing inhomogeneous dielectrics},
  author = {Dung, Ho Trung and Kn{\"o}ll, Ludwig and Welsch, Dirk-Gunnar},
  journal = {Phys. Rev. A},
  volume = {57},
  pages = {3931--3942},
  year = {1998},
  doi = {10.1103/PhysRevA.57.3931}
}

@article{Franke2019,
  title = {Quantization of quasinormal modes for open cavities and plasmonic cavity quantum electrodynamics},
  author = {Franke, Sebastian and Hughes, Stephen and Dezfouli, Mohsen Kamandar and Kristensen, Philip Tr{\o}st and Busch, Kurt and Knorr, Andreas and Richter, Marten},
  journal = {Phys. Rev. Lett.},
  volume = {122},
  pages = {213901},
  year = {2019},
  doi = {10.1103/PhysRevLett.122.213901}
}

@article{Gustin2025Reservoir,
  title = {What is the spectral density of the reservoir for a lossy quantized cavity?},
  author = {Gustin, Chris and Ren, Juanjuan and Hughes, Stephen},
  journal = {Phys. Rev. Lett.},
  volume = {134},
  pages = {123601},
  year = {2025},
  doi = {10.1103/PhysRevLett.134.123601}
}

@article{Gustin2023,
  title = {Gauge-invariant theory of truncated quantum light-matter interactions in arbitrary media},
  author = {Gustin, Chris and Franke, Sebastian and Hughes, Stephen},
  journal = {Phys. Rev. A},
  volume = {107},
  pages = {013722},
  year = {2023},
  doi = {10.1103/PhysRevA.107.013722},
}

@article{GustinErratum2025,
  title = {Erratum: Gauge-invariant theory of truncated quantum light-matter interactions in arbitrary media [{P}hys. {R}ev. {A} 107, 013722 (2023)]},
  author = {Gustin, Chris and Franke, Sebastian and Hughes, Stephen},
  journal = {Phys. Rev. A},
  volume = {112},
  pages = {039904},
  year = {2025},
  doi = {10.1103/zwzq-4k6z},
}

@article{Scala2021,
  title = {Beyond the {R}abi model: Light interactions with polar atomic systems in a cavity},
  author = {Scala, Giovanni and S{\l}owik, Karolina and Facchi, Paolo and Pascazio, Saverio and Pepe, Francesco V.},
  journal = {Phys. Rev. A},
  volume = {104},
  pages = {013722},
  year = {2021},
  doi = {10.1103/PhysRevA.104.013722},
}

@article{Gross2018,
  title = {Near-field strong coupling of single quantum dots},
  author = {Gro{\ss}, Heiko and Hamm, Joachim M. and Tufarelli, Tommaso and Hess, Ortwin and Hecht, Bert},
  journal = {Sci. Adv.},
  volume = {4},
  pages = {eaar4906},
  year = {2018},
  doi = {10.1126/sciadv.aar4906},
}

@article{Hu2024,
  title = {Robust consistent single quantum dot strong coupling in plasmonic nanocavities},
  author = {Hu, Shu and Huang, Junyang and Arul, Rakesh and S{\'a}nchez-Iglesias, Ana and Xiong, Yuling and Liz-Marz{\'a}n, Luis M. and Baumberg, Jeremy J.},
  journal = {Nat. Commun.},
  volume = {15},
  pages = {6835},
  year = {2024},
  doi = {10.1038/s41467-024-51170-7},
}

@article{Fry2000,
  title = {Inverted electron-hole alignment in {InAs}-{GaAs} self-assembled quantum dots},
  author = {Fry, P. W. and Itskevich, I. E. and Mowbray, D. J. and Skolnick, M. S. and Finley, J. J. and Barker, J. A. and O'Reilly, E. P. and Wilson, L. R. and Larkin, I. A. and Maksym, P. A. and Hopkinson, M. and Al-Khafaji, M. and David, J. P. R. and Cullis, A. G. and Hill, G. and Clark, J. C.},
  journal = {Phys. Rev. Lett.},
  volume = {84},
  pages = {733--736},
  year = {2000},
  doi = {10.1103/PhysRevLett.84.733},
}

@article{Huo2014,
  title = {A light-hole exciton in a quantum dot},
  author = {Huo, Y. H. and Witek, B. J. and Kumar, S. and Cardenas, J. R. and Zhang, J. X. and Akopian, N. and Singh, R. and Zallo, E. and Grifone, R. and Kriegner, D. and Trotta, R. and Ding, F. and Stangl, J. and Zwiller, V. and Bester, G. and Rastelli, A. and Schmidt, O. G.},
  journal = {Nat. Phys.},
  volume = {10},
  pages = {46--51},
  year = {2014},
  doi = {10.1038/nphys2799},
}

@article{Huang2021,
  title = {Electric field induced tuning of electronic correlation in weakly confining quantum dots},
  author = {Huang, Huiying and Csontosov{\'a}, Diana and Manna, Santanu and Huo, Yongheng and Trotta, Rinaldo and Rastelli, Armando and Klenovsk{\'y}, Petr},
  journal = {Phys. Rev. B},
  volume = {104},
  pages = {165401},
  year = {2021},
  doi = {10.1103/PhysRevB.104.165401},
}

@article{Vurgaftman2001,
  title = {Band parameters for {III-V} compound semiconductors and their alloys},
  author = {Vurgaftman, I. and Meyer, J. R. and Ram-Mohan, L. R.},
  journal = {J. Appl. Phys.},
  volume = {89},
  pages = {5815--5875},
  year = {2001},
  doi = {10.1063/1.1368156},
}

@article{Sambe1973,
  title = {Steady states and quasienergies of a quantum-mechanical system in an oscillating field},
  author = {Sambe, Hideo},
  journal = {Phys. Rev. A},
  volume = {7},
  pages = {2203--2213},
  year = {1973},
  doi = {10.1103/PhysRevA.7.2203},
}

@article{FriskKockum2019,
  title = {Ultrastrong coupling between light and matter},
  author = {Frisk Kockum, Anton and Miranowicz, Adam and De Liberato, Simone and Savasta, Salvatore and Nori, Franco},
  journal = {Nat. Rev. Phys.},
  volume = {1},
  pages = {19--40},
  year = {2019},
  doi = {10.1038/s42254-018-0006-2},
}

@article{delValle2012,
  title = {Theory of frequency-filtered and time-resolved {N}-photon correlations},
  author = {del Valle, E. and Gonzalez-Tudela, A. and Laussy, F. P. and Tejedor, C. and Hartmann, M. J.},
  journal = {Phys. Rev. Lett.},
  volume = {109},
  pages = {183601},
  year = {2012},
  doi = {10.1103/PhysRevLett.109.183601}
}

@article{delValle2016Erratum,
  title = {Erratum: Theory of frequency-filtered and time-resolved {N}-photon correlations [{Phys. Rev. Lett.} {109}, 183601 (2012)]},
  author = {del Valle, E. and Gonzalez-Tudela, A. and Laussy, F. P. and Tejedor, C. and Hartmann, M. J.},
  journal = {Phys. Rev. Lett.},
  volume = {116},
  pages = {249902},
  year = {2016},
  doi = {10.1103/PhysRevLett.116.249902}
}

@article{Reiter2019,
  title = {Distinctive characteristics of carrier-phonon interactions in optically driven semiconductor quantum dots},
  author = {Reiter, D. E. and Kuhn, T. and Axt, V. M.},
  journal = {Adv. Phys. X},
  volume = {4},
  pages = {1655478},
  year = {2019},
  doi = {10.1080/23746149.2019.1655478},
}

@misc{reproduction-package,
  title = {{A} semiconductor photon-pair source based on a polariton cascade: reproduction package},
  author = {Kawa, Karol},
  year = {2026},
  howpublished = {Zenodo},
  note = {Version 1.0, \href{https://doi.org/10.5281/zenodo.23476265}{doi:10.5281/zenodo.23476265}},
  doi = {10.5281/zenodo.23476265},
}

\end{document}